\documentclass[a4paper,12pt]{article}
\pdfoutput=1 

\usepackage{graphicx}
\usepackage{amsmath}
\usepackage{amssymb}
\usepackage{amsfonts}
\usepackage{amsthm}
\usepackage{stmaryrd} 
\usepackage{xcolor}
\definecolor{refkey}{rgb}{0.40, 0.55, 0.55}
\definecolor{labelkey}{rgb}{0.40, 0.55, 0.55}
\usepackage{cite}
\newcommand{\del}{\partial}

 \makeatletter
 \@addtoreset{equation}{section}
 \makeatother

\usepackage[bookmarks=true,bookmarksnumbered=true,bookmarkstype=toc]{hyperref} 
\hypersetup{
pdfauthor={Tetsuji KIMURA; Shin Sasaki; Kenta Shiozawa},
colorlinks={true},
linkcolor={black},
urlcolor={blue},
filecolor={black},
citecolor={blue}
}
\usepackage{cancel}
\usepackage{ulem}

\newcommand{\dop}{\mathrm{d}}

\newcommand{\TS}{\hspace{.5pt}} 
\usepackage{mathrsfs} 

\date{empty}
\begin{document}
\allowdisplaybreaks{
\begin{titlepage}
\null
\begin{flushright}
March,
2022
\end{flushright}
\vskip 2cm
\begin{center}
{\Large \bf 
Complex Structures, T-duality and 
\\
\vskip 0.1cm
Worldsheet Instantons in Born Sigma Models 
}
\vskip 2cm
\normalsize
\renewcommand\thefootnote{\alph{footnote}}

{\large
Tetsuji Kimura${}^{*}$\footnote{t-kimura(at)osakac.ac.jp}, 
Shin Sasaki${}^{\dagger}$\footnote{shin-s(at)kitasato-u.ac.jp}
and 
Kenta Shiozawa${}^{\dagger}$\footnote{k.shiozawa(at)sci.kitasato-u.ac.jp}
}

\vskip 0.5cm

  {\it
  ${}^{*}$
  Center for Physics and Mathematics, Institute for Liberal Arts and
 Sciences, \\
  Osaka Electro-Communication University, Neyagawa, Osaka 572-8530,
 Japan \\
  \vspace{0.3cm}
  ${}^{\dagger}$
  Department of Physics, Kitasato University \\
  Sagamihara 252-0373, Japan
  }
\vskip 1cm
\begin{abstract}
We investigate doubled (generalized) complex structures in
 $2D$-dimensional Born geometries where T-duality symmetry is manifestly realized.
We show that K\"{a}hler, hyperk\"{a}hler, bi-hermitian and bi-hypercomplex
 structures of spacetime are implemented in Born geometries as doubled structures.
We find that the Born structures and the generalized K\"{a}hler
 (hyperk\"{a}hler) structures appear as subalgebras of bi-quaternions and
split-tetra-quaternions.
We find parts of these structures are classified by Clifford algebras.
We then study the T-duality nature of the worldsheet instantons in Born
 sigma models. 
We show that the instantons in K\"{a}hler geometries are
 related to those in bi-hermitian geometries in a non-trivial way.

\end{abstract}
\end{center}

\end{titlepage}

\newpage
\setcounter{footnote}{0}
\renewcommand\thefootnote{\arabic{footnote}}
\pagenumbering{arabic}
\tableofcontents

\section{Introduction} \label{sect:introduction}
One of the important features that characterizes string theories is
duality \cite{Obers:1998fb}.
T-duality, that distinguishes string theory from theories based on
point particles, is the most distinctive feature to understand stringy nature of
spacetime.

T-duality among spacetime geometries 
is studied in various contexts.
For example, the famous Buscher rule of T-duality \cite{Buscher:1987sk}
is derived in the two-dimensional string sigma model as a target space transformation.
In supersymmetric theories, a duality symmetry between 
chiral and twisted chiral multiplets of two-dimensional $\mathcal{N} = (2,2)$ sigma models is
interpreted as T-duality \cite{Lindstrom:1983rt, Gates:1983nr, Rocek:1991ps}.
In general, an ${\cal N}=(2,2)$ theory only with chiral multiplets requires that the target
space geometry is K\"{a}hler \cite{Zumino:1979et, Alvarez-Gaume:1980xat}.
In particular, the presence of the twisted chiral multiplets
requires that the target space is the bi-hermitian geometry admitting two independent complex structures
$(J_+, J_-)$ that 
commute with each other and 
are compatible with the 
target space metric \cite{Gates:1984nk, Howe:1984fak}.
A pair of noncommuting complex structures in $\mathcal{N} = (2,2)$
models with semi-chiral multiplets is also studied \cite{Goteman:2009xb}.
It is shown in the sigma model language that K\"{a}hler and
bi-hermitian geometries are T-dual with each other \cite{Ivanov:1994ec,
Hassan:1994mq, Bakas:1995hc, Hassan:1995je}.
Similarly, $\mathcal{N} = (4,4)$ 
supersymmetry requires that the target space is generically 
a bi-hypercomplex geometry which admits
three sets of complex structures 
$(J_{a,+}, J_{a,-})$ (with $a=1,2,3$)
satisfying $J_{a,+} J_{b,-} = J_{b,-} J_{a,+}$.

On the other hand, geometric realization of T-duality symmetry is developed in
the context of generalized geometry \cite{Hitchin:2003, Gualtieri:2004}.
The generalized tangent bundle $\mathbb{T}M$ over a $D$-dimensional spacetime manifold $M$ is defined
by the formal sum of the tangent and the cotangent bundles 
$\mathbb{T}M = TM \oplus T^*M$. 
The spacetime geometry $M$ is encoded by $2D$-dimensional generalized
structures on $\mathbb{T}M$ in the $O(D,D)$ covariant fashion.
For example, the K\"{a}hler structure on spacetime $M$ is realized as
the generalized K\"{a}hler structure $(\mathcal{J}_J, \mathcal{J}_{\omega})$ 
on $\mathbb{T}M$, via so-called the Gualtieri map \cite{Gualtieri:2004}.
It is also shown that the bi-hermitian structure 
on $M$ is
realized by the generalized K\"{a}hler structure $(\mathcal{J}_{+}, \mathcal{J}_{-})$.
The physical origin of this correspondence is studied in 
supersymmetric sigma models \cite{Lindstrom:2004eh, Lindstrom:2004iw,
Bredthauer:2005zx, Lindstrom:2005zr, Zabzine:2005qf, Bredthauer:2006hf}.
This also holds for the hyperk\"{a}hler and the bi-hypercomplex cases.
They are realized as the generalized hyperk\"{a}hler structures on $\mathbb{T}M$.

Generalized geometry is closely related to doubled formalism
\cite{Duff:1989tf, Siegel:1993th, Siegel:1993xq}.
The idea of T-duality symmetric geometries is further sophisticated in
the study of double field theory (DFT) \cite{Hull:2009mi}.
DFT is developed on the basis of the doubled formalism in which the spacetime metric $g_{\mu \nu}$, the
NSNS $B$-field and the 
dilaton $\phi$ are organized into the $2D \times 2D$ 
generalized metric $\mathcal{H}_{MN}$ and the generalized 
dilaton $d$.
They are defined in the $2D$-dimensional doubled space $\mathcal{M}$
where T-duality symmetry is manifestly and geometrically realized.
T-duality symmetries of geometric quantities are implemented as global $O(D,D)$ transformations in
the doubled space.
For example, the Buscher rule of $g_{\mu \nu}$, $B_{\mu \nu}$ and $\phi$ is reproduced by an $O(D,D)$
rotation of the generalized metric $\mathcal{H}_{MN}$ and the generalized 
dilaton $d$.
The general T-duality transformation law of K\"{a}hler, hyperk\"{a}hler, bi-hermitian
and bi-hypercomplex structures of spacetime geometries are also discussed in
the doubled formalism \cite{Kimura:2022dma}.
The geometry of the doubled space $\mathcal{M}$ is implemented by the Born
structures \cite{Vaisman:2012ke, Freidel:2013zga, Freidel:2017yuv,
Freidel:2018tkj, Marotta:2018myj}.
The Born geometry is endowed with the doubled foliations, the $O(D,D)$
structure, the natural inner product by the $O(D,D)$ invariant metric $\eta_{MN}$, 
the generalized metric $\mathcal{H}_{MN}$ and a unique connection.
Furthermore, it is shown that the tangent bundle of the doubled space
$T\mathcal{M}$ is identified with the generalized tangent bundle
$\mathbb{T}M$ through a natural isomorphism.

The purpose of this paper is to study the T-duality nature of 
the spacetime structures of K\"{a}hler, hyperk\"{a}hler, bi-hermitian and bi-hypercomplex
geometries by embedding them into extensions of the generalized
(hyper)K\"{a}hler structures and the Born structures. We call these doubled structures in general.
In particular, we study 
compatibility of the doubled structures with
the Born geometry on which DFT is naturally defined.
We will show that the Born geometry is compatible with the generalized
(hyper)k\"{a}hler structures by encoding the complex structures of
spacetime into doubled structures in an appropriate way. 
Along the way, we will encounter interesting connections between the doubled
structures and certain algebras of hypercomplex numbers.
We will analyze the algebras that the doubled structures obey.
With these results, in the latter half of this paper, 
we study the T-duality covariant expression of the worldsheet instantons.
The existence of the generalized complex structures in the doubled space leads us to the
notion of doubled worldsheet instantons.
We study the doubled worldsheet instantons in the Born sigma model which is a
sigma model whose target space is the $2D$-dimensional Born geometry \cite{Marotta:2019eqc}.
This provides us a T-duality covariant way of string worldsheet theory.

The organization of this paper is as follows.
In the next section, we introduce the Born geometry on which DFT is
defined. The relation between generalized geometry and doubled geometry
is discussed.
In section \ref{sec:generalized_complex}, we study the 
compatibility of the Born and the generalized K\"{a}hler structures.
We study algebras of hypercomplex numbers that these structures obey.
We show that K\"{a}hler, hyperk\"{a}hler, bi-hermitian and
bi-hypercomplex structures of spacetime are embedded into doubled
structures together with the Born geometry.
In section \ref{sec:Born_sigma_model}, we study the T-duality covariant
expression of the worldsheet instanton equations in the Born sigma model.
We discuss T-duality property of the worldsheet instantons.
Section \ref{sec:conclusion} is devoted to conclusion and discussions.
Appendix \ref{sec:appendix1} and \ref{sec:appendix2} are glossaries of mathematics on hypercomplex
numbers and Clifford algebras.

\section{Double field theory and Born geometry} \label{sec:Born_geometry}
In this section, we clarify the relations among double field theory,
the Born geometry and generalized geometry.

\subsection{Double field theory}
We start by introducing double field theory (DFT) \cite{Hull:2009mi}.
DFT is a formulation of supergravities for which T-duality is manifestly
realized. The fundamental fields of DFT are the generalised metric
$\mathcal{H}_{MN}$ and the generalized 
dilaton $d$.
They are defined in the $2D$-dimensional doubled space $\mathcal{M}$.
The doubled coordinate $\mathbb{X}^M , \, (M=1, \ldots, 2D)$ on
$\mathcal{M}$ is decomposed as $\mathbb{X}^M = (X^{\mu},
\tilde{X}_{\mu}),  \, (\mu = 1, \ldots, D)$ where $X^{\mu}$ and
$\tilde{X}_{\mu}$ are the Kaluza-Klein (KK) and the winding 
coordinates respectively. The action of DFT is given by
\begin{align}
S_{\text{DFT}} = \int \! {\dop}^{2D} \mathbb{X} \, 
e^{-2d}
\mathscr{R} (\mathcal{H},d),
\label{eq:DFT_action}
\end{align}
where $\mathscr{R} (\mathcal{H}, d)$ is the generalized Ricci scalar
defined by
\begin{align}
\mathscr{R} (\mathcal{H},d) =& \ 
4 \mathcal{H}^{MN} \del_M \del_N d
- \del_M \del_N \mathcal{H}^{MN}
- 4 \mathcal{H}^{MN} \del_M d \del_N d
+ 4 \del_M \mathcal{H}^{MN} \del_N d
\notag \\
& 
+ \frac{1}{8} \mathcal{H}^{MN} \del_M \mathcal{H}^{KL} \del_N
 \mathcal{H}_{KL}
- \frac{1}{2} \mathcal{H}^{MN} \del_M \mathcal{H}^{KL} \del_K \mathcal{H}_{NL}.
\end{align}
Here the derivative means $\del_M = \frac{\del}{\del \mathbb{X}^M}$. 
The indices are raised and lowered by the $O(D,D)$ invariant metric
$\eta_{MN}$ and its inverse $\eta^{MN}$.
The action \eqref{eq:DFT_action} is manifestly invariant under the
global $O(D,D)$ transformation and is invariant under
the $O(D,D)$ covariantized diffeomorphism and 
the $B$-field gauge transformation.
The T-duality transformations of 
the spacetime fields are implemented by the $O(D,D)$ rotation
of the generalized metric $\mathcal{H}_{MN} (\mathbb{X})$ and the generalized
dilaton $d (\mathbb{X})$.
To see this, it is useful to employ the standard parameterization of the DFT
quantities
\begin{align}
&
\mathcal{H}_{MN} = 
\left(
\begin{array}{cc}
g_{\mu \nu} - B_{\mu \rho} g^{\rho \sigma} B_{\sigma \nu} & B_{\mu \rho} g^{\rho \nu}
 \\
- g^{\mu \rho} B_{\rho \nu} & g^{\mu \nu}
\end{array}
\right)
, \qquad 
e^{-2d} = \sqrt{-g} e^{-2\phi},
\notag \\
&
\eta_{MN} = 
\begin{pmatrix}
 0 & \delta_{\mu} {}^{\nu} \\
 \delta^{\mu} {}_{\nu} & 0
\end{pmatrix}
, \qquad
\eta^{MN} =
\begin{pmatrix}
0 & \delta^{\mu} {}_{\nu} \\
\delta_{\mu} {}^{\nu} & 0
\end{pmatrix}
,
\label{eq:DFT_parametrization}
\end{align}
where 
$g_{\mu \nu}$ and $B_{\mu \nu}$ are $D \times D$ symmetric and
anti-symmetric matrices respectively,
while 
$\phi$ is a real function on $\mathcal{M}$.
The generalized metric $\mathcal{H}_{MN}$ is an $O(D,D)$ element and $d$ is invariant under the
$O(D,D)$ rotation. All the quantities involving the gauge parameters in DFT are subject to the constraints
\begin{align}
\del_M \del^M * = 0, \qquad \del_M * \del^M * = 0,
\label{eq:DFT_constraints}
\end{align}
where $*$ are any quantities in DFT.
The first equation in \eqref{eq:DFT_constraints} is the level-matching condition of closed strings 
while the second one is specific to DFT. This is called the strong constraint.

The physical spacetime $M$ is defined by a $D$-dimensional slice in the
doubled space $\mathcal{M}$. This is defined by a solution to the
constraints \eqref{eq:DFT_constraints}.
For example, when all the components in $\mathcal{H}_{MN}$, $d$ and
gauge parameters depend only on $X^{\mu}$, the constraints
\eqref{eq:DFT_constraints} are trivially satisfied.
In this case, a slice $\tilde{X}_{\mu} = \mathrm{const}.$, parameterized
by $X^{\mu}$, is chosen as 
the $D$-dimensional physical spacetime.
The components $g_{\mu \nu} (X)$, $B_{\mu \nu} (X)$ and $\phi (X)$ are
identified with the spacetime metric, the NSNS $B$-field and the dilaton, respectively.
Then, the action \eqref{eq:DFT_action} reduces to that of the NSNS
sector of type II supergravities
\begin{align}
S = \int \! {\dop}^D X \, \sqrt{-g} e^{-2\phi} 
\Bigg[
R + 4 (\del \phi)^2 - \frac{1}{12} (H^{(3)})^2
\Bigg],
\end{align}
where $R$ is the Ricci scalar defined by the spacetime metric
$g_{\mu \nu}$ and $H^{(3)} = {\dop} B$ is the field strength of the
$B$-field.
In this sense, DFT is a T-duality covariant formulation of supergravity.
 
We next move to a more sophisticated treatment of the doubled space
$\mathcal{M}$ and discuss its geometric structures.

\subsection{Born geometry}
The structures of the doubled space $\mathcal{M}$ 
in the previous subsection are furnished in the Born manifold
\cite{Vaisman:2012ke, Freidel:2013zga, Freidel:2017yuv,
Freidel:2018tkj}.
Before introducing the Born manifold, we start from 
an almost para-complex manifold.
Given a $2D$-dimensional differential manifold $\mathcal{M}$, 
an endomorphism $\mathcal{K}: T \mathcal{M} \to T \mathcal{M}$ that
satisfies $\mathcal{K}^2 = \mathbf{1}_{2D}$ 
and whose $\pm 1$-eigenbundle has the same rank
is called an almost para-complex structure.
Then the pair $(\mathcal{M},\mathcal{K})$ 
gives an almost para-complex manifold. 
Since $\mathcal{K}$ is a real analogue of the almost complex structure
$J^2 = - 1$, we call this kind endomorphism the almost real structure.
We also call endomorphisms on $T\mathcal{M}$ the doubled structures in general.
Due to the property $\mathcal{K}^2 = \mathbf{1}_{2D}$, we have two eigenbundles
 (distributions) $L$ and $\tilde{L}$ in $T\mathcal{M}$ associated with
 two eigenvalues $\mathcal{K} = \pm 1$.
They are defined by the projection operators $P = \frac{1}{2} (\mathbf{1}_{2D} +
\mathcal{K})$, $\tilde{P} = \frac{1}{2} (\mathbf{1}_{2D} - \mathcal{K})$
as $L = P(T\mathcal{M})$, $\tilde{L} = \tilde{P} (T \mathcal{M})$.
Then the tangent space of the
almost para-complex manifold $\mathcal{M}$
is decomposed as $T \mathcal{M} = L \oplus \tilde{L}$.

We then introduce the notion of integrability of doubled structures.
For a distribution $\mathcal{D} \subset T \mathcal{M}$ and 
vector fields $X,Y \in \Gamma (\mathcal{D})$, when the Lie bracket 
$\llbracket \cdot, \cdot \rrbracket$
evaluated on $X$ and $Y$ becomes again a vector field in $\mathcal{D}$,
namely 
$\llbracket X, Y \rrbracket \in \Gamma(\mathcal{D})$, then $\mathcal{D}$ is called involutive.
By the Frobenius theorem, a distribution $\mathcal{D}$ is integrable if and only if it is involutive.
With this definition, we consider the projected Lie brackets,
\begin{align}
N_P (X,Y) = \tilde{P} 
\llbracket P(X), P(Y) \rrbracket, \qquad 
N_{\tilde{P}} (X,Y) = P 
\llbracket \tilde{P} (X), \tilde{P} (Y) \rrbracket.
\end{align}
Apparently $N_P = 0$ implies the involutivity and hence the integrability of $L$.
The same is true for $N_{\tilde{P}} = 0$ and the integrability of $\tilde{L}$.
The Nijenhuis tensor of the endomorphism ${\cal K}$ is defined by 
\begin{align}
N_{\mathcal{K}} (X,Y) = N_P (X,Y) + N_{\tilde{P}} (X,Y),
\end{align}
where $X,Y \in \Gamma (T\mathcal{M})$.
Since $N_{\mathcal{K}} = 0$ means that $\mathcal{K}$ is integrable, it
is obvious that the integrabilities of $L$ and $\tilde{L}$ imply that of $\mathcal{K}$.
An almost para-complex manifold $(\mathcal{M},\mathcal{K})$ is said to
be a para-complex manifold when $\mathcal{K}$ is integrable.
On the other hand, when only $L (\tilde{L})$ is integrable, 
$(\mathcal{M}, \mathcal{K})$ is an $L (\tilde{L})$-para-complex manifold.

We next introduce a metric in 
the almost para-complex manifold
$(\mathcal{M}, \mathcal{K})$.
A neutral metric $\eta$ of signature $(D,D)$ is defined by a map 
$\eta : T \mathcal{M} \times T \mathcal{M} \to \mathbb{R}$.
When this satisfies 
$\eta (\mathcal{K} \cdot, \mathcal{K} \cdot) = - \eta (\cdot, \cdot)$,
it is called a para-hermitian metric.
The triple $(\mathcal{M}, \mathcal{K}, \eta)$ is called an almost para-hermitian manifold.
Although, the para-hermitian metric $\eta_{MN}$ is not necessarily
flat, we always take it to be flat in this paper.
In this case, the metric $\eta_{MN}$ is identified with the $O(D,D)$ invariant metric in DFT.
By the neutral metric $\eta_{MN}$ and $\mathcal{K}^M {}_N$, we define the
fundamental 
two-form $(\omega_{\mathcal{K}})_{MN} = \eta_{ML} \mathcal{K}^L {}_N$ 
which is not closed in general. 
When $\omega_{\mathcal{K}}$ is non-degenerate, it defines an
almost symplectic structure on $\mathcal{M}$.
Then the para-hermitian structure defines an almost symplectic manifold $(\mathcal{M}, \omega_{\mathcal{K}})$.
When ${\dop} \omega_{\mathcal{K}} = 0$, then $\omega_{\mathcal{K}}$ is a
symplectic structure on $\mathcal{M}$ by which the
non-degenerate Poisson structure
$\{\cdot, \cdot\}_{\text{P}} = \omega_{\mathcal{K}}^{-1} ({\dop}\cdot, {\dop}\cdot)$
is induced.
In this case $\mathcal{M}$ is an almost para-K\"{a}hler manifold.
When $\mathcal{K}$ is integrable, $(\mathcal{M}, \omega_{\mathcal{K}})$
is a para-K\"{a}hler manifold. This is also known as a bi-Lagrangian manifold.
The almost para-hermitian manifolds are classified by 
the integrability of $\mathcal{K}$ and the closedness of $\omega_{\mathcal{K}}$.
See Table \ref{tb:almost_para}.
\begin{table}[t]
\centering
\begin{tabular}{c|c|c}
 & ${\dop} \omega_{\mathcal{K}} \not= 0$ & ${\dop} \omega_{\mathcal{K}} = 0$ \\
\hline
$N_{\mathcal{K}} \not= 0$ & almost para-hermitian & almost para-K\"{a}hler \\
\hline
$N_{\mathcal{K}} = 0$ & para-hermitian & para-K\"{a}hler
\end{tabular}
 \caption{The classification of almost para-hermitian manifolds
 $(\mathcal{M}, \mathcal{K})$.}
\label{tb:almost_para}
\end{table}

The last quantity we introduce is the metric $\mathcal{H}_{MN}$ of signature $(2D,0)$.
Let $(\mathcal{M},\mathcal{K},\eta)$ be a para-hermitian manifold.
We define $\mathcal{H}$ as a Riemannian metric of signature $(2D,0)$
that satisfies
\begin{align}
\eta^{-1} \mathcal{H} = \mathcal{H}^{-1} \eta, \qquad 
\omega^{-1}_{\mathcal{K}} \mathcal{H} = - \mathcal{H}^{-1} \omega_{\mathcal{K}}.
\label{eq:eta_H_compatibility}
\end{align}
Then $(\eta, \omega_{\mathcal{K}}, \mathcal{H})$ 
is called the Born structure 
and the quadruple $(\mathcal{M}, \eta, \omega_{\mathcal{K}}, \mathcal{H})$
is the Born manifold.
In DFT language, $\mathcal{H}_{MN}$ is identified with the generalized metric.
The condition \eqref{eq:eta_H_compatibility} is rewritten as 
\begin{align}
(\eta^{-1} \mathcal{H})^2 = \mathbf{1}_{2D}, \qquad
(\mathcal{H}^{-1} \omega_{\mathcal{K}})^2 = - \mathbf{1}_{2D}.
\end{align}
This means that the quantity $\mathcal{J} = \eta^{-1} \mathcal{H}$ defines an almost real
structure $\mathcal{J}^2 = \mathbf{1}_{2D}$ on $T\mathcal{M}$ and the compatibility condition becomes 
\begin{align}
\eta (\mathcal{J} \cdot, \mathcal{J} \cdot) = \eta (\cdot,\cdot).
\end{align}
The pair $(\eta, \mathcal{J})$ is called the chiral structure on $\mathcal{M}$.
On the other hand, the quantity $\mathcal{I} = \mathcal{H}^{-1} \omega_{\mathcal{K}}$
defines an almost complex structure $\mathcal{I}^2 = - \mathbf{1}_{2D}$ on $T\mathcal{M}$ and the condition becomes 
\begin{align}
\mathcal{H} (\mathcal{I} \cdot, \mathcal{I} \cdot) = \mathcal{H} (\cdot,\cdot).
\end{align}
The pair $(\mathcal{H},\mathcal{I})$ is called an almost hermitian
structure on $\mathcal{M}$.
The equations \eqref{eq:eta_H_compatibility} are compatibility conditions
on $\eta$, $\mathcal{H}$ and $\omega_{\mathcal{K}}$.
Since $\mathcal{K} = \eta^{-1} \omega_{\mathcal{K}}$ satisfies $\mathcal{K}^2
= \mathbf{1}_{2D}$, the condition becomes
\begin{align}
\omega_{\mathcal{K}} (\mathcal{K} \cdot, \mathcal{K} \cdot) = - \omega_{\mathcal{K}} (\cdot,\cdot).
\end{align}

In summary, a $2D$-dimensional Born manifold $\mathcal{M}$ is equipped with 
the neutral metric $\eta$ of signature $(D,D)$, 
the fundamental two-form $\omega_{\mathcal{K}}$ 
and the Riemannian metric $\mathcal{H}$ of signature $(2D,0)$. 
The pair $(\omega_{\mathcal{K}},\mathcal{K})$ is 
the para-hermitian structure, 
$(\eta,\mathcal{J})$ is the chiral structure, 
$(\mathcal{H},\mathcal{I})$ is the almost hermitian structure on $\mathcal{M}$.
The triple $(\eta,\omega_{\mathcal{K}},\mathcal{H})$ is the Born
structure which defines $\mathcal{I} = \mathcal{H}^{-1} \omega_{\mathcal{K}} = - \omega^{-1}_{\mathcal{K}} \mathcal{H}$, 
$\mathcal{J} = \eta^{-1} \mathcal{H} = \mathcal{H}^{-1} \eta$, $\mathcal{K} =
 \eta^{-1} \omega_{\mathcal{K}} = \omega^{-1}_{\mathcal{K}} \eta$.
The doubled structure $(\mathcal{I},\mathcal{J},\mathcal{K})$ is called an almost
para-quaternionic structure on $\mathcal{M}$ and satisfies
\begin{gather}
- \mathcal{I}^2 = \mathcal{J}^2 = \mathcal{K}^2 = \mathbf{1}_{2D},
 \qquad \mathcal{IJK} = - \mathbf{1}_{2D},
\notag \\
\{ \mathcal{I},\mathcal{J} \} = \{\mathcal{J},\mathcal{K}\} =
 \{\mathcal{K},\mathcal{I}\} = 0.
\label{eq:para-quaternion}
\end{gather}
Here $\{\cdot, \cdot\}$ is the anti-commutator of the doubled structures.
The consistency conditions for Born structure are summarized as follows;
\begin{gather}
\begin{alignat}{3}
\mathcal{H} (\mathcal{I} X,\mathcal{I} Y) &= \mathcal{H} (X,Y), 
&\qquad 
\mathcal{H} (\mathcal{J} X,\mathcal{J} Y) &= \mathcal{H} (X,Y),
&\qquad 
\mathcal{H} (\mathcal{K} X,\mathcal{K} Y) &= \mathcal{H} (X,Y),
\notag \\
\eta (\mathcal{I} X, \mathcal{I} Y) &= - \eta (X,Y),
&\qquad 
\eta (\mathcal{J} X, \mathcal{J} Y) &= \eta (X,Y), 
&\qquad 
\eta (\mathcal{K} X,\mathcal{K} Y) &= - \eta (X,Y),
\notag \\
\omega_{\mathcal{K}} (\mathcal{I} X,\mathcal{I} Y) &= \omega_{\mathcal{K}} (X,Y),
&\qquad 
\omega_{\mathcal{K}} (\mathcal{J} X,\mathcal{J} X) &= - \omega_{\mathcal{K}} (X,Y),
&\qquad 
\omega_{\mathcal{K}} (\mathcal{K} X,\mathcal{K} Y) &= - \omega_{\mathcal{K}} (X,Y),
\notag 
\end{alignat}
\\
X,Y \in \Gamma (T\mathcal{M}).
\end{gather}
Altogether we call these structures the Born geometry.

\subsection{Born geometry and generalized geometry}
The para-hermitian structure $(\omega_{\mathcal{K}},\mathcal{K})$ of a Born manifold defines 
the eigenbundles $L$ and $\tilde{L}$.
By the Frobenius theorem, the involutive bundle $L$ defines a
foliation structure in $\mathcal{M}$ that allows $L = T\mathcal{F}$.
The physical spacetime in DFT is identified with a leaf of the foliation.
When we write the basis of $L$ as $\del_{\mu} = \frac{\del}{\del X^{\mu}}$, 
then the local coordinate of the base leaves $\mathcal{F}$ is $X^{\mu}$.
The same is true for $\tilde{L}$. 
Since $\tilde{L}$ is integrable in the para-hermitian manifold, there is a foliation structure that
defines leaves whose coordinate is $\tilde{X}_{\mu}$, and the basis of
$\tilde{L}$ is given by $\tilde{\del}^{\mu} = \frac{\del}{\del \tilde{X}_{\mu}}$.
The pair $(\mathcal{F}, \tilde{\mathcal{F}})$ defines a doubled
foliation in $\mathcal{M}$ and we find the natural coordinate
system $\mathbb{X}^M = (X^{\mu},\tilde{X}_{\mu})$ in $\mathcal{M}$.
This is identified with the KK and the winding coordinates.
The $D$-dimensional physical spacetime $M$ in a $2D$-dimensional Born
manifold $\mathcal{M}$ is a leaf in $\mathcal{F}$ or $\tilde{\mathcal{F}}$.
A physical field $\Phi$ on $\mathcal{M}$ is given by an (anti)para-holomorphic
quantity $\tilde{{\dop}} \Phi = 0 \ ({\dop} \Phi = 0)$ defined by $\mathcal{K}$ \cite{Mori:2019slw}.
This is a trivial solution to the constraints \eqref{eq:DFT_constraints}.

We now examine the relation between the doubled space and the generalized tangent bundle on a $D$-dimensional
physical space $M$.
The generalized tangent bundle of $M$ is the formal sum of the tangent
and the cotangent bundles $\mathbb{T}
M = TM \oplus T^*M$.
Since we have the neutral metric $\eta$ in $\mathcal{M}$, there is a map $T\mathcal{M} = L \oplus \tilde{L} \to L^*
\oplus \tilde{L}^*$. This induces the following isomorphisms;
\begin{align}
\phi^+ : \tilde{L} \to L^*, \qquad \phi^- : L \to \tilde{L}^*.
\end{align}
Then $\tilde{L}$ is identified with the dual vector space $L^*$ of $L$.
This defines natural isomorphisms \cite{Vaisman:2012ke, Freidel:2017yuv, Freidel:2018tkj};
\begin{align}
\Phi^+ : T \mathcal{M} \to L \oplus L^*, \qquad \Phi^- : T \mathcal{M}
 \to \tilde{L} \oplus \tilde{L}^*.
\label{eq:natural_isomorphism}
\end{align}
The distribution $L = T \mathcal{F}$ is the tangent
bundle of the leaves $\mathcal{F}$ and $L^*$ is its dual $T^*\mathcal{F}$.
Therefore $T\mathcal{M}$ is identified with the generalized tangent
bundle $T \mathcal{F} \oplus T^* \mathcal{F}$ over $\mathcal{F}$, or $T
\tilde{\mathcal{F}} \oplus T^* \tilde{\mathcal{F}}$ over
$\tilde{\mathcal{F}}$.
In the following, we choose $M \subset \mathcal{F}$ without loss of
generality and identify doubled structures on $T\mathcal{M}$ with
generalized structures on $\mathbb{T}M$ through the natural isomorphism \eqref{eq:natural_isomorphism}.
In this case, doubled vectors and generalized vectors are identified as follows;
\begin{align}
V = V^M \del_M = V^{\mu} \del_{\mu} + \tilde{V}_{\mu} \tilde{\del}^{\mu} 
\qquad 
\Longleftrightarrow
\qquad 
V = V^{\mu} \del_{\mu} + \tilde{V}_{\mu} {\dop} X^{\mu}.
\end{align}

For later convenience, we introduce a particular representation of the Born structure;
\begin{gather}
\mathcal{H}_{MN} = 
\left(
\begin{array}{cc}
g_{\mu \nu} - B_{\mu \rho} g^{\rho \sigma} B_{\sigma \nu} & B_{\mu \rho} g^{\rho \nu}
 \\
- g^{\mu \rho} B_{\rho \nu} & g^{\mu \nu}
\end{array}
\right), 
\qquad
\eta_{MN} = 
\left(
\begin{array}{cc}
0 & \delta_{\mu} {}^{\nu}
 \\
\delta^{\mu} {}_{\nu} & 0
\end{array}
\right)
, 
\notag \\
(\omega_{\mathcal{K}})_{MN}
=
\left(
\begin{array}{cc}
2 B_{\mu \nu} & - \delta_{\mu} {}^{\nu}
 \\
\delta^{\mu} {}_{\nu} & 0
\end{array}
\right)
.
\label{eq:Born_structure_representation1}
\end{gather}
Note that when we impose the constraints \eqref{eq:DFT_constraints} in
DFT, $g_{\mu \nu}$ and $B$ are the spacetime metric and the $B$-field on
a leaf in $\mathcal{F}$.
The other structures are parameterized as 
\begin{align}
\mathcal{I}^M {}_N =& \ 
\mathcal{H}^{ML} (\omega_{\mathcal{K}})_{LN}
=
\left(
\begin{array}{cc}
g^{\mu \rho} B_{\rho \nu} & - g^{\mu \nu}
\\
 g_{\mu \nu} + B_{\mu \rho} g^{\rho \sigma} B_{\sigma \nu} & - B_{\mu
 \rho} g^{\rho \nu}
\end{array}
\right),
\notag \\
\mathcal{J}^M {}_N =& \ \eta^{ML} \mathcal{H}_{LN} = 
\left(
\begin{array}{cc}
- g^{\mu \rho} B_{\rho \nu} & g^{\mu \nu}
\\ 
g_{\mu \nu} - B_{\mu \rho} g^{\rho \sigma} B_{\sigma \nu} & B_{\mu \rho}
 g^{\rho \nu}
\end{array}
\right),
\notag \\
\mathcal{K}^M {}_N =& \ \eta^{ML} (\omega_{\mathcal{K}})_{LN} = 
\left(
\begin{array}{cc}
\delta^{\mu} {}_{\nu} & 0
\\ 
2 B_{\mu \nu} & - \delta_{\mu} {}^{\nu}
\end{array}
\right).
\label{eq:Born_structure_representation2}
\end{align}
With this parameterization, 
we find that
$\mathcal{I}$, $\mathcal{J}$ and $\mathcal{K}$ anti-commute with each
other and they indeed satisfy
\begin{align}
\mathcal{I}^M {}_L \mathcal{I}^L {}_N
= - \delta^M {}_N,
\qquad
\mathcal{J}^M {}_L \mathcal{J}^L {}_N 
= \delta^M {}_N,
\qquad
\mathcal{K}^M {}_L \mathcal{K}^L {}_N
 = \delta^M {}_N.
\end{align}
We call the equations \eqref{eq:Born_structure_representation1}, \eqref{eq:Born_structure_representation2} the standard representation.

We note that the neutral metric $\eta$ defines a natural inner product
on the doubled or generalized vectors $U$ and $V$;
\begin{align}
\langle U, V \rangle = \eta^{MN} V_M U_N = v^{\mu} \tilde{u}_{\mu} +
 \tilde{v}_{\mu} u^{\mu},
\end{align}
where we have used the expansions 
$U = u^{\mu} \del_{\mu} + \tilde{u}_{\mu} \tilde{\del}^{\mu}$ 
and 
$V = v^{\mu} \del_{\mu} + \tilde{v}_{\mu} \tilde{\del}^{\mu}$.

\section{Born structure and generalized complex structures} \label{sec:generalized_complex}

In this section, we study the compatibility conditions for the Born
structure on $\mathcal{M}$ and the K\"{a}hler structure on $M$.
The K\"{a}hler structure on spacetime $M$ is embedded into
the generalized K\"{a}hler structure on $\mathbb{T}M$ via the Gualtieri
map. This is identified with a doubled structure on $T \mathcal{M}$ via
the natural isomorphism. 
We analyze algebras that govern the Born geometry and the generalized
K\"{a}hler structures.
We also examine how they are combined into the doubled space $\mathcal{M}$.

\subsection{Embedding K\"{a}hler structures}
It is widely known that an (almost) complex structure $J$ in spacetime $M$ is embedded into a
generalized almost complex structure $\mathcal{J}$ on $\mathbb{T}M$ \cite{Gualtieri:2004}.
A generalized almost complex structure is an endomorphism
$\mathcal{J} : \mathbb{T}M \to \mathbb{T} M$ that preserves the inner
product 
$\langle \mathcal{J} \cdot, \mathcal{J} \cdot \rangle = \langle \cdot, \cdot \rangle$ 
and squares to minus identity 
$\mathcal{J}^2 = - \mathbf{1}_{2D}$. 
Since $\mathcal{J}^2 = - \mathbf{1}_{2D}$, the generalized almost complex structure defines
$\pm i$-eigenbundles on the complexified generalized tangent bundle;
\begin{align}
l_{\pm} = 
\Big\{
V \in \mathbb{T} M \otimes \mathbb{C} : \mathcal{J} V = \pm i V
\Big\}.
\end{align}
The eigenbundles $l_{\pm}$ have the complex rank $D$ and are maximally isotropic and $l_+ \cap l_- = 0$. 
The integrability of the generalized almost complex structures are
defined through the Courant involutivity of the eigenbundles \cite{Gualtieri:2004}. 
When this is the case, $\mathcal{J}$ becomes
a generalized complex structure.
A generalized K\"{a}hler structure is defined by a pair of two commuting
generalized complex structures $(\mathcal{J}_1, \mathcal{J}_2)$ whose
product $\mathcal{G} = \mathcal{J}_1 \mathcal{J}_2$ defines a
positive-definite metric on $\mathbb{T} M$.

With these definitions, we exhibit an explicit example of the generalized
K\"{a}hler structure.
Given a complex structure $J$ in a K\"{a}hler manifold $M$, we
have generalized complex structures of the form;
\begin{align}
\mathcal{J}_J 
= 
\left(
\begin{array}{cc}
J & 0  \\
0 & - J^*
\end{array}
\right),
\qquad 
\mathcal{J}_{\omega}
=
\left(
\begin{array}{cc}
0 & - \omega^{-1} \\
\omega & 0
\end{array}
\right),
\label{eq:generalized_complex}
\end{align}
where $J^*$ is the adjoint of $J$ and $\omega = - g J$ is the K\"{a}hler
two-form associated with the complex structure $J$.
It is shown that $\mathcal{J}_J$ and $\mathcal{J}_{\omega}$ are Courant
involutive when $J$ is integrable and ${\dop} \omega = 0$ which holds for
any K\"{a}hler manifold $M$.
We find that $\mathcal{J}_J$ and $\mathcal{J}_{\omega}$ commute with
each other and their product 
\begin{align}
\mathcal{G} 
= 
\mathcal{J}_J \mathcal{J}_{\omega}
= 
\left(
\begin{array}{cc}
0 & - J \omega^{-1} \\
- J^* \omega & 0
\end{array}
\right)
= 
\left(
\begin{array}{cc}
0 & g^{-1}  \\
g & 0
\end{array}
\right)
\end{align}
becomes a positive-definite metric on $\mathbb{T}M$
\footnote{
Unless otherwise stated, we consider the Euclidean metric $g_{\mu \nu}$
in the following.
}.
Then the pair $(\mathcal{J}_J, \mathcal{J}_{\omega})$ defines a generalized K\"{a}hler structure.

We examine the compatibility of the generalized K\"{a}hler structure
$(\mathcal{J}_J, \mathcal{J}_{\omega})$ and the Born structure on $\mathcal{M}$.
Since the physical spacetime $M \subset \mathcal{F}$ admits the
metric $g_{\mu \nu}$ and the K\"{a}hler structure, namely, an integrable complex
structure $J$ and a symplectic form $\omega = - g J$, 
they satisfy $J^2 = -1$ and $\omega J = - J^* \omega$ on $T\mathcal{F} = L$.
We also assume $B=0$ for the time being.
In the standard representation, the 
building blocks of the doubled structure
$\mathcal{I}$, $\mathcal{J}$ and $\mathcal{K}$ 
in the Born manifold are given by
\begin{align}
\mathcal{I} = 
\left(
\begin{array}{cc}
0 & - g^{-1}  \\
g & 0
\end{array}
\right),
\quad
\mathcal{J} = 
\left(
\begin{array}{cc}
0 & g^{-1} \\
g & 0
\end{array}
\right),
\quad
\mathcal{K} = 
\left(
\begin{array}{cc}
1 & 0  \\
0 & -1
\end{array}
\right).
\label{eq:Born_representation_B_zero}
\end{align}
They obey the algebra of the almost para-quaternionic structure
\eqref{eq:para-quaternion}.
In other words, this is the algebra of the split-quaternions involving
two real and one imaginary units (see Appendix \ref{sec:appendix1}).
On the other hand, the generalized K\"{a}hler structure 
$(\mathcal{J}_J, \mathcal{J}_{\omega}, \mathcal{G})$
obeys the algebra
\begin{gather}
- \mathcal{J}_J^2 
= - \mathcal{J}_{\omega}^2 
= \mathcal{G}^2 
= \mathbf{1}_{2D}, \qquad 
\mathcal{J}_{J} \mathcal{J}_{\omega} \mathcal{G} 
= \mathbf{1}_{2D},
\notag \\
[\mathcal{J}_J, \mathcal{J}_{\omega}] 
= [\mathcal{J}_{\omega}, \mathcal{G}] 
= [\mathcal{G}, \mathcal{J}_J] 
= 0.
\end{gather}
Here $[\cdot, \cdot]$ is the commutator of the doubled structures.
This is the algebra of the bi-complex numbers $\mathbb{C}_2$ (see
Appendix \ref{sec:appendix1}).
We examine algebraic structures that incorporate the split-quaternions
and the bi-complex numbers as subalgebras.
By the explicit calculation, it is obvious that the almost product structure $\mathcal{G}$ in the generalized K\"{a}hler
structure gives the chiral structure in the Born structure, $\mathcal{G}
= \mathcal{J}$.
Hereafter we denote $\mathcal{J}$ instead of $\mathcal{G}$.
The products of the generalized K\"{a}hler structure 
$(\mathcal{J}_J, \mathcal{J}_{\omega})$
and the doubled structure $(\mathcal{I}, \mathcal{J},
\mathcal{K})$ introduce the additional structures on $T\mathcal{M}$;
\begin{align}
\mathcal{J}_J \mathcal{I} 
&=
\mathcal{I} \mathcal{J}_J 
=
\left(
\begin{array}{cc}
0 & - J g^{-1} \\
- J^* g & 0
\end{array}
\right)
=
\left(
\begin{array}{cc}
0 & - \omega^{-1} \\
- \omega & 0
\end{array}
\right) = \mathcal{P},
\notag \\
\mathcal{J}_J \mathcal{K} 
&=
\mathcal{K} \mathcal{J}_J
=
\left(
\begin{array}{cc}
J & 0 \\
0 & J^*
\end{array}
\right)
=
\mathcal{Q},
\notag \\
\mathcal{J}_{\omega} \mathcal{I} 
&=
- \mathcal{I} \mathcal{J}_{\omega}
=
\left(
\begin{array}{cc}
- \omega^{-1} g & 0  \\
0 & - \omega g^{-1}
\end{array}
\right)
=
\left(
\begin{array}{cc}
- J & 0 \\
0 & - J^*
\end{array}
\right) = 
- \mathcal{Q},
\notag \\ 
\mathcal{J}_{\omega} \mathcal{K} 
&= 
- \mathcal{K} \mathcal{J}_{\omega}
=
\left(
\begin{array}{cc}
0 & \omega^{-1} \\
\omega & 0
\end{array}
\right) = - \mathcal{P}.
\label{eq:LM_definition}
\end{align}
This means that the algebra is not closed by 
$(\mathcal{J}_J, \mathcal{J}_{\omega}, \mathcal{I}, \mathcal{J}, \mathcal{K})$.
The newly appeared structures $\mathcal{P}$ and $\mathcal{Q}$ 
satisfy
\begin{align}
\mathcal{P}^2 = \mathbf{1}_{2D}, \qquad \mathcal{Q}^2 = - \mathbf{1}_{2D}.
\end{align}
They play as real and complex structures on $T \mathcal{M}$.
By evaluating all the products involving $\mathcal{P}$ and $\mathcal{Q}$, we find
\begin{alignat}{3}
\mathcal{J}_J \mathcal{P} &= \mathcal{P} \mathcal{J}_J = - \mathcal{I}, 
&\qquad
\mathcal{J}_{\omega} \mathcal{P} &= - \mathcal{P} \mathcal{J}_{\omega} = \mathcal{K},
&\qquad
\mathcal{IP} &= \mathcal{PI} = - \mathcal{J}_J,
\notag \\
\mathcal{JP} &= - \mathcal{PJ} = \mathcal{Q},
&\qquad
\mathcal{KP} &= - \mathcal{PK} = \mathcal{J}_{\omega},
&\qquad
\mathcal{QP} &= - \mathcal{PQ} = \mathcal{J},
\notag \\
\mathcal{J}_J \mathcal{Q} &= \mathcal{Q} \mathcal{J}_J = - \mathcal{K},
&\qquad
\mathcal{J}_{\omega} \mathcal{Q} &= - \mathcal{Q} \mathcal{J}_{\omega} = \mathcal{I},
&\qquad
\mathcal{IQ} &= - \mathcal{QI} = - \mathcal{J}_{\omega},
\notag \\
\mathcal{JQ} &= - \mathcal{QJ} = \mathcal{P},
&\qquad
\mathcal{KQ} &= \mathcal{QK} = \mathcal{J}_J.
\label{eq:LM_products}
\end{alignat}
Here we have used the relation $\omega = - g J$ in the evaluation of the
products.
From \eqref{eq:LM_products}, we find that no additional structures appear.
Then the algebra is closed by the basis 
\begin{align}
(\mathbf{1}_{2D},\mathcal{J}_J, \mathcal{J}_{\omega},
\mathcal{I}, \mathcal{J},\mathcal{K}, \mathcal{P}, \mathcal{Q}).
\end{align}
In this algebra, we have four complex and four real structures on $T \mathcal{M}$;
\begin{gather}
\mathcal{J}_J^2 = \mathcal{J}_{\omega}^2 = \mathcal{I}^2 = \mathcal{Q}^2
 = - \mathbf{1}_{2D},
\notag \\
\mathbf{1}_{2D}^2 = \mathcal{J}^2 = \mathcal{K}^2 = \mathcal{P}^2 = \mathbf{1}_{2D}.
\end{gather}
The basis 
$(\mathbf{1}_{2D},\mathcal{J}_J, \mathcal{J}_{\omega}, \mathcal{I}, \mathcal{J},\mathcal{K}, \mathcal{P}, \mathcal{Q})$ 
defines an eight-dimensional algebra whose product table
is given in Table \ref{tb:product_table_gkb}.
\begin{table}[t]
\centering
\begin{tabular}{c||c|c|c|c|c|c|c|c}
  & $\mathbf{1}_{2D}$ & $\mathcal{J}$ & $\mathcal{K}$ & $\mathcal{P}$ 
  & $\mathcal{J}_J$ & $\mathcal{J}_{\omega}$ & $\mathcal{I}$ & $\mathcal{Q}$ \\
\hline
\hline
$\mathbf{1}_{2D}$ 
  & $\mathbf{1}_{2D}$ & $\mathcal{J}$ & $\mathcal{K}$ & $\mathcal{P}$ 
  & $\mathcal{J}_J$ & $\mathcal{J}_{\omega}$ & $\mathcal{I}$ & $\mathcal{Q}$ \\
\hline
$\mathcal{J}$ 
  & $\mathcal{J}$ & $\mathbf{1}_{2D}$ & $\mathcal{I}$ & $\mathcal{Q}$ 
  & $-\mathcal{J}_{\omega}$ & $-\mathcal{J}_J$ &  $\mathcal{K}$ & $\mathcal{P}$ \\
\hline
$\mathcal{K}$ 
  & $-\mathcal{K}$ & $-\mathcal{I}$ & $\mathbf{1}_{2D}$ & $\mathcal{J}_{\omega}$ 
  & $\mathcal{Q}$ & $\mathcal{P}$ & $-\mathcal{J}$ & $\mathcal{J}_J$ \\
\hline
$\mathcal{P}$ 
  & $\mathcal{P}$ & $-\mathcal{Q}$ & $-\mathcal{J}_{\omega}$ & $\mathbf{1}_{2D}$ 
  & $-\mathcal{I}$ & $-\mathcal{K}$ & $-\mathcal{J}_J$ & $-\mathcal{J}$ \\
\hline
$\mathcal{J}_J$ 
  & $\mathcal{J}_J$ & $-\mathcal{J}_{\omega}$ & $\mathcal{Q}$ & $-\mathcal{I}$ 
  & $-\mathbf{1}_{2D}$ & $\mathcal{J}$ & $\mathcal{P}$ & $-\mathcal{K}$ \\
\hline
$\mathcal{J}_{\omega}$ 
  & $\mathcal{J}_{\omega}$ & $-\mathcal{J}_J$ & $-\mathcal{P}$ & $\mathcal{K}$ 
  & $\mathcal{J}$ & $-\mathbf{1}_{2D}$ & $-\mathcal{Q}$ & $\mathcal{I}$ \\
\hline
$\mathcal{I}$ 
  & $\mathcal{I}$ & $-\mathcal{K}$ & $\mathcal{J}$ & $-\mathcal{J}_J$ 
  & $\mathcal{P}$ & $\mathcal{Q}$ & $-\mathbf{1}_{2D}$ & $-\mathcal{J}_{\omega}$ \\
\hline
$\mathcal{Q}$ 
  & $\mathcal{Q}$ & $-\mathcal{P}$ & $-\mathcal{J}_J$ & $\mathcal{J}$ 
  & $-\mathcal{K}$ & $-\mathcal{I}$ & $\mathcal{J}_{\omega}$ & $-\mathbf{1}_{2D}$ \\
\end{tabular}
\caption{The product table including the Born and the generalized
 complex structures. Left $\times$ right products are shown.}
\label{tb:product_table_gkb}
\end{table}
We find that this is the algebra of the bi-quaternions.
It is known that algebras of some hypercomplex numbers are isomorphic to
 Clifford algebras.
Indeed, the bi-quaternion algebras are equivalent to the Clifford
algebras $Cl_{3,0} (\mathbb{R})$, $Cl_{2,1} (\mathbb{R})$ and 
$Cl_{1,2}(\mathbb{R})$ (see Appendix \ref{sec:appendix2}).

As we show in Appendix \ref{sec:appendix1}, there are commutative and anti-commutative bases 
which form subalgebras in the bi-quaternion algebra.
The subalgebras include bi-complex numbers $\mathbb{C}_2$, 
split-quaternions $\mathrm{Sp}\mathbb{H}$ and
quaternions $\mathbb{H}$.
We find all the subalgebras of the bi-quaternion algebra;
\begin{quote}
\begin{enumerate}
\item \label{enum:3.1-1}
$\mathbb{C}_2$ \ : \ 
$(\mathbf{1}_{2D},\mathcal{J}_J, \mathcal{J}_{\omega}, \mathcal{J})$,
$\mathcal{J}_J^2 = \mathcal{J}_{\omega}^2 = -\mathbf{1}_{2D}$,
$\mathcal{J}^2 = \mathbf{1}_{2D}$\ ; commutative,
\item \label{enum:3.1-2}
$\mathbb{C}_2$ \ : \ 
$(\mathbf{1}_{2D},\mathcal{J}_J, \mathcal{I}, \mathcal{P})$,
$\mathcal{J}_J^2 = \mathcal{I}^2 = -\mathbf{1}_{2D}$, 
$\mathcal{P}^2 = \mathbf{1}_{2D}$\ ; commutative,
\item \label{enum:3.1-3}
$\mathbb{C}_2$ \ : \ 
$(\mathbf{1}_{2D},\mathcal{J}_J, \mathcal{K}, \mathcal{Q})$,
$\mathcal{J}_J^2 = \mathcal{Q}^2 = -\mathbf{1}_{2D}$, 
$\mathcal{K}^2 = \mathbf{1}_{2D}$\ ; commutative,
\item \label{enum:3.1-4}
$\mathrm{Sp}\mathbb{H}$ \ : \ 
$(\mathbf{1}_{2D},\mathcal{I}, \mathcal{J}, \mathcal{K})$, 
$\mathcal{I}^2 = -\mathbf{1}_{2D}$, 
$\mathcal{J}^2 = \mathcal{K}^2 = \mathbf{1}_{2D}$\ ; anti-commutative,
\item \label{enum:3.1-5}
$\mathrm{Sp}\mathbb{H}$ \ : \ 
$(\mathbf{1}_{2D},\mathcal{J}, \mathcal{P}, \mathcal{Q})$, 
$\mathcal{Q}^2 = -\mathbf{1}_{2D}$, 
$\mathcal{J}^2 = \mathcal{P}^2 = \mathbf{1}_{2D}$\ ; anti-commutative,
\item \label{enum:3.1-6}
$\mathrm{Sp}\mathbb{H}$ \ : \ 
$(\mathbf{1}_{2D},\mathcal{J}_{\omega}, \mathcal{K}, \mathcal{P})$, 
$\mathcal{J}_{\omega}^2 = -\mathbf{1}_{2D}$, 
$\mathcal{K}^2 = \mathcal{P}^2 = \mathbf{1}_{2D}$\ ; anti-commutative,
\item \label{enum:3.1-7}
$\mathbb{H}$ \ : \ 
$(\mathbf{1}_{2D},\mathcal{J}_{\omega}, \mathcal{I}, \mathcal{Q})$,
$\mathcal{J}_{\omega}^2 = \mathcal{Q}^2 = \mathcal{I}^2 = -\mathbf{1}_{2D}$\ ; anti-commutative.
\end{enumerate}
\end{quote}
Note that the algebra of split-quaternions 
(\ref{enum:3.1-4}), (\ref{enum:3.1-5}) and (\ref{enum:3.1-6}),
defining the Born structure,
is isomorphic to Clifford algebras $\mathrm{Sp}\mathbb{H} \simeq
Cl_{2,0} (\mathbb{R}) \simeq Cl_{1,1} (\mathbb{R})$. 
The quaternions (\ref{enum:3.1-7}) defines a hypercomplex structure on the doubled
space $\mathcal{M}$ whose realization in Clifford algebra is $Cl_{0,2} (\mathbb{R})$.
The bi-complex numbers (\ref{enum:3.1-1}), (\ref{enum:3.1-2}) and (\ref{enum:3.1-3}), 
in Clifford language $Cl_1 (\mathbb{C})$,
define the generalize K\"{a}hler structures.

We then examine the compatibility conditions of the structures
$(\mathcal{J}_J, \mathcal{J}_{\omega}, \mathcal{I},  
\mathcal{J}, \mathcal{K}, \mathcal{P}, \mathcal{Q})$ and the metrics 
$\eta$, $\mathcal{H}$ on $\mathcal{M}$.
The building blocks of the Born structures $\mathcal{I}$, $\mathcal{J}$ and $\mathcal{K}$ satisfy 
\begin{align}
\eta(\mathcal{I} \cdot, \mathcal{I} \cdot) = - \eta (\cdot, \cdot),
\qquad
\eta(\mathcal{J} \cdot, \mathcal{J} \cdot) =   \eta (\cdot, \cdot),
\qquad
\eta(\mathcal{K} \cdot, \mathcal{K} \cdot) = - \eta (\cdot, \cdot).
\label{eq:compatibility1}
\end{align}
Since $\eta$ defines the natural inner product on 
$T\mathcal{M} \simeq \mathbb{T}M$, the generalized K\"{a}hler structure 
$(\mathcal{J}_J, \mathcal{J}_{\omega})$ satisfies
\begin{align}
\eta(\mathcal{J}_J \cdot, \mathcal{J}_J \cdot) = \eta (\cdot, \cdot),
\qquad 
\eta(\mathcal{J}_{\omega} \cdot, \mathcal{J}_{\omega} \cdot) = \eta (\cdot, \cdot).
\end{align}
Note that this together with 
$\mathcal{J} = \mathcal{J}_J \mathcal{J}_{\omega}$
implies the second condition in \eqref{eq:compatibility1}.
Furthermore, since $\mathcal{P} = \mathcal{K} \mathcal{J}_{\omega}$ and 
$\mathcal{Q} = - \mathcal{J}_{\omega} \mathcal{I}$, we have 
\begin{gather}
\eta (\mathcal{P} \cdot, \mathcal{P} \cdot)
=
\eta (\mathcal{K} \mathcal{J}_{\omega} \cdot, \mathcal{K} \mathcal{J}_{\omega} \cdot)
=
- 
\eta (\mathcal{J}_{\omega} \cdot, \mathcal{J}_{\omega} \cdot)
=
- \eta (\cdot, \cdot),
\notag \\
\eta (\mathcal{Q} \cdot, \mathcal{Q} \cdot)
=
\eta (\mathcal{J}_{\omega} \mathcal{I} \cdot, \mathcal{J}_{\omega} \mathcal{I} \cdot)
= \eta (\mathcal{I} \cdot, \mathcal{I} \cdot)
= - \eta (\cdot, \cdot).
\end{gather}
This means that $\eta$ is anti-hermitian with respect to 
$\mathcal{P}$ and $\mathcal{Q}$.

We have the compatibility conditions for the Born structures;
\begin{align}
\mathcal{H} (\mathcal{I} \cdot, \mathcal{I} \cdot) = \mathcal{H} (\cdot, \cdot),
\qquad
\mathcal{H} (\mathcal{J} \cdot, \mathcal{J} \cdot) = \mathcal{H} (\cdot,\cdot),
\qquad
\mathcal{H} (\mathcal{K} \cdot, \mathcal{K} \cdot) = \mathcal{H} (\cdot, \cdot).
\end{align}
The metric of the K\"{a}hler spacetime $M$ satisfies the hermitian
condition $g (J \cdot, J \cdot) = g (\cdot, \cdot)$. This implies the following properties;
\begin{align}
\mathcal{H} (\mathcal{J}_J \cdot, \mathcal{J}_J \cdot) 
= \mathcal{H} (\cdot, \cdot),
\qquad 
\mathcal{H} (\mathcal{J}_{\omega} \cdot, \mathcal{J}_{\omega} \cdot) = \mathcal{H} (\cdot, \cdot).
\end{align}
Then, the compatibility conditions for $\mathcal{P}$ and $\mathcal{Q}$
on the generalized metric $\mathcal{H}$ is found to be
\begin{align}
\mathcal{H} (\mathcal{P} \cdot, \mathcal{P} \cdot) 
&= 
\mathcal{H} (\mathcal{K} \mathcal{J}_{\omega} \cdot, \mathcal{K} \mathcal{J}_{\omega} \cdot)
=
\mathcal{H} (\cdot, \cdot),
\notag \\
\mathcal{H} (\mathcal{Q} \cdot, \mathcal{Q} \cdot)
&=
\mathcal{H} (\mathcal{J}_{\omega} \mathcal{I} \cdot, \mathcal{J}_{\omega} \mathcal{I} \cdot)
=
\mathcal{H} (\cdot, \cdot).
\end{align}

We here comment on the effects of the $B$-field on the above structures.
Until now, we have not cared about the $B$-field.
For a given (almost) real or complex structure $\mathcal{A}$ on $T \mathcal{M}$, 
the $B$-field is introduced by an $O(D,D)$ transformation on $T \mathcal{M}$, known as the $B$-transformation
\begin{align}
\mathcal{A}^B = e^B \mathcal{A} e^{-B}, \qquad 
e^B = 
\left(
\begin{array}{cc}
1 & 0 \\
B & 1
\end{array}
\right).
\end{align}
Indeed, the equation \eqref{eq:Born_representation_B_zero} 
becomes the standard representation
\eqref{eq:Born_structure_representation2} by the $B$-transformation.
The same is true for the representation \eqref{eq:LM_definition}.
Since the $B$-transformation is a similarity transformation, the algebra
closes even in the presence of the $B$-field.

In summary, the K\"{a}hler structure $(J,\omega)$
of spacetime $M$ is embedded into the
doubled structures on $T \mathcal{M}$ satisfying the algebra of bi-quaternions.
The compatibility of the generalized K\"{a}hler and the Born structures
requires the bi-quaternion algebra that encompasses  
four real and four complex structures 
$(\mathbf{1}_{2D}, \mathcal{J}_J, \mathcal{J}_{\omega}, \mathcal{I}, \mathcal{J}, \mathcal{K}, \mathcal{P}, \mathcal{Q})$.
The algebra has substructures given by bi-complex numbers, 
split-quaternions and quaternions.

\subsection{Embedding bi-hermitian structures}
We next consider embeddings of the bi-hermitian structure 
$(J_{\pm}, \omega_{\pm})$ of spacetime $M$. 
Note that the metric $g_{\mu \nu}$ is hermitian with
respect to $J_{\pm}$, and $\omega_{\pm} = - g J_{\pm}$ are the
fundamental two-forms.
The bi-hermitian structures are embedded into the generalized complex
structures $\mathcal{J}_{\pm}$ as \cite{Gualtieri:2004};
\begin{align}
\mathcal{J}_{\pm} = \frac{1}{2} 
\left(
\mathcal{J}_{J_+} \pm \mathcal{J}_{J_-} + \mathcal{J}_{\omega_+} \mp \mathcal{J}_{\omega_-}
\right),
\end{align}
where the matrices of $\mathcal{J}_{J_{\pm}}$ and
$\mathcal{J}_{\omega_{\pm}}$ are given by \eqref{eq:generalized_complex}.
We find that $\mathcal{J}_+$ and $\mathcal{J}_-$ commute with each other and
they give the chiral structure $\mathcal{J} = \mathcal{J}_+ \mathcal{J}_-$.
The triples $(\mathcal{J}_{J_{\pm}}, \mathcal{J}_{\omega_{\pm}}, \mathcal{J})$ 
form the algebras of two bi-complex numbers sharing the
common real (chiral) structure $\mathcal{J}$.
Since the algebra of the bi-quaternions does not support such subalgebras, 
we need to enlarge the algebra to incorporate the bi-hermitian structure 
$(J_{\pm}, \omega_{\pm})$ into the Born geometry.

We find that an algebra that allows subalgebras of two bi-complex numbers sharing one real structure $\mathcal{J}$
is the bi-quaternions over the field $\mathbb{C}$. 
This is a 16-dimensional algebra and schematically written as
$\mathbb{C} \otimes \mathbb{C} \otimes \mathbb{H}$, 
which is isomorphic to the Clifford algebra $Cl_3 (\mathbb{C})$.
This algebra contains 8 real and 8 imaginary units and the bi-complex
numbers are subalgebras generated by the bases (see Appendix \ref{sec:appendix1});
\begin{align}
(e_0 \mathbf{i} \hat{1}, e_i \mathbf{1} \hat{1}, e_i \mathbf{i}
 \hat{1}),
\qquad 
(e_0 \mathbf{1} \hat{i}, e_i \mathbf{i} \hat{i}, e_i \mathbf{i}
 \hat{1}),
\qquad
(i=1,2,3).
\label{eq:bh_bi-complex}
\end{align}
Here $e_{\mu}$, $(\mathbf{1}, \mathbf{i})$ and $(\hat{1}, \hat{i})$ are
bases of $\mathbb{H}$, $\mathbb{C}$ and $\mathbb{C}$, respectively.
The bases \eqref{eq:bh_bi-complex} share
the real unit $e_i \mathbf{i} \hat{1}$.
For example, if we assign 
\begin{align}
\mathcal{J} = e_1 \mathbf{i} \hat{1},
\quad
\mathcal{J}_{J_+} 
= e_0 \mathbf{i} \hat{1},
\quad
\mathcal{J}_{\omega_+} = e_1 \mathbf{1} \hat{1},
\quad
\mathcal{J}_{J_-} 
= e_0 \mathbf{1} \hat{i},
\quad
\mathcal{J}_{\omega_-} = - e_1 \mathbf{i} \hat{i},
\end{align}
we find that they obey the algebras of two bi-complex numbers.
The Born structure $(\mathcal{I}, \mathcal{J}, \mathcal{K})$ is
represented by a split-quaternion subalgebra of $\mathbb{C} \otimes
\mathbb{C} \otimes \mathbb{H}$. This is given by the basis
\begin{align}
(e_1 \mathbf{i} \hat{1}, e_2 \mathbf{i} \hat{1}, e_3 \mathbf{1} \hat{1}).
\end{align}
Therefore we employ
the following assignment;
\begin{align}
\mathcal{I} = e_3 \mathbf{1} \hat{1},
\qquad
\mathcal{J} = e_1 \mathbf{i} \hat{1},
\qquad
\mathcal{K} = e_2 \mathbf{i} \hat{1}.
\end{align}
We have five imaginary units 
$\mathcal{J}_{J_{\pm}}, \mathcal{J}_{\omega_{\pm}}, \mathcal{I}$.
The other three imaginary units 
$\mathcal{P}', \mathcal{Q}', \mathcal{R}'$
are represented as
\begin{align}
\mathcal{P}' = e_2 \mathbf{1} \hat{1},
\qquad
\mathcal{Q}' = e_2 \mathbf{i} \hat{i},
\qquad
\mathcal{R}' = e_3 \mathbf{i} \hat{i}.
\end{align}
Since we have relations
\begin{align}
e_2 \mathbf{1} \hat{1} = (e_3 \mathbf{1} \hat{1}) (e_1 \mathbf{1}
 \hat{1}),
\qquad
e_2 \mathbf{i} \hat{i} = (e_3 \mathbf{1} \hat{1}) (e_1 \mathbf{i}
 \hat{i}),
\qquad
e_3 \mathbf{i} \hat{i} = (e_3 \mathbf{1} \hat{1}) (e_0 \mathbf{1}
 \hat{i}) (e_0 \mathbf{i} \hat{1}),
\end{align}
$\mathcal{P}', \mathcal{Q}', \mathcal{R}'$
are obtained as (without the $B$-field)
\begin{align}
\mathcal{P}' 
&= \mathcal{I} \mathcal{J}_{\omega_+}
= 
\left(
\begin{array}{cc}
0 & - g^{-1} \\
g & 0
\end{array}
\right)
\left(
\begin{array}{cc}
0 & - \omega_+^{-1} \\
\omega_+ & 0 
\end{array}
\right)
=
-
\left(
\begin{array}{cc}
J_+ & 0  \\
0 & J^*_+
\end{array}
\right),
\notag \\
\mathcal{Q}' 
&= 
- \mathcal{I} \mathcal{J}_{\omega_-}
=
\left(
\begin{array}{cc}
0 & - g^{-1} \\
g & 0
\end{array}
\right)
\left(
\begin{array}{cc}
0 & - \omega_-^{-1} \\
\omega_- & 0 
\end{array}
\right)
=
\left(
\begin{array}{cc}
J_- & 0  \\
0 & J^*_-
\end{array}
\right),
\notag \\
\mathcal{R}' 
&=  
\mathcal{I} \mathcal{J}_{J_-} \mathcal{J}_{J_+}
=
\left(
\begin{array}{cc}
0 & - g^{-1} \\
g & 0
\end{array}
\right)
\left(
\begin{array}{cc}
J_- & 0  \\
0 & - J^*_-
\end{array}
\right)
\left(
\begin{array}{cc}
J_+ & 0  \\
0 & - J^*_+
\end{array}
\right)
=
\left(
\begin{array}{cc}
0 & - g^{-1} J^*_- J^*_+ \\
g J_- J_+ & 0
\end{array}
\right).
\end{align}
We easily confirm that $\mathcal{P}'{}^2 = \mathcal{Q}'{}^2 = - \mathbf{1}_{2D}$ and 
\begin{align}
\mathcal{R}'{}^2 = 
\begin{pmatrix}
- g^{-1} J^*_- J^*_+ g J_- J_+ & 0  \\
0 & - g J_- J_+ g^{-1} J^*_- J^*_+
\end{pmatrix}
=
\begin{pmatrix}
- J_- J_+ J_- J_+ & 0  \\
0 & - g J_- J_+ J_- J_+ g^{-1}
\end{pmatrix}
= - \mathbf{1}_{2D}.
\end{align}
Here we have used the relations of the bi-hermitian structure 
$J^*_{\pm} = - g J_{\pm} g^{-1}$
and the fact that $J_+$ and $J_-$ commute with each other.

Similarly, the additional real units 
$(\mathcal{S}', \mathcal{T}', \mathcal{U}', \mathcal{V}', \mathcal{W}')$
other than 
$(\mathbf{1}_{2D},\mathcal{J}, \mathcal{K})$ 
are found to be 
\begin{alignat}{2}
\mathcal{S}' 
&=  e_3 \mathbf{i} \hat{1} = \mathcal{I} \mathcal{J}_{J_+}
 = 
-
\begin{pmatrix}
0 &  \omega_+^{-1} \\
 \omega_+ & 0
\end{pmatrix},
&\qquad
\mathcal{T}' 
&= e_1 \mathbf{1} \hat{i} = \mathcal{J}_{\omega_+}
 \mathcal{J}_{J_-} = 
\begin{pmatrix}
0 & \omega_+^{-1} J_-^* \\
\omega_+ J_- & 0
\end{pmatrix},
\notag \\
\mathcal{U}' 
&= e_2 \mathbf{1} \hat{i} 
= \mathcal{P}' \mathcal{J}_{J_-}
 = 
\begin{pmatrix}
- J_+ J_- & 0 \\
0 & J_+^* J_-^*
\end{pmatrix},
&\qquad
\mathcal{V}' 
&= e_3 \mathbf{1} \hat{i} = \mathcal{I} \mathcal{J}_{J_-}
 = -
\begin{pmatrix}
 0 & \omega_-^{-1} \\
\omega_- & 0
\end{pmatrix},
\notag \\
\mathcal{W}' 
&= e_0 \mathbf{i} \hat{i} = \mathcal{J}_{J_+}
 \mathcal{J}_{J_-}
= 
\begin{pmatrix}
J_+ J_- & 0 \\
0 & J_+^* J_-^*
\end{pmatrix}.
\end{alignat}
Note that the subalgebra by 
$(\mathbf{1}_{2D}, \mathcal{J}_{J_{\pm}}, \mathcal{J}_{\omega_{\pm}}, \mathcal{J}, \mathcal{T}', \mathcal{W}')$
involving the generalized K\"{a}hler structure 
$(\mathcal{J}_{J_{\pm}}, \mathcal{J}_{\omega_{\pm}})$
forms the algebra of the tri-complex
numbers $\mathbb{C}_3$ elucidated in \cite{Kimura:2022dma}.

\subsection{Embedding hyperk\"{a}hler and bi-hypercomplex structures}
The hyperk\"{a}hler structure $(J_a, \omega_a)$ $(a=1,2,3)$ on $M$ is embedded into
the generalized hyperk\"{a}hler structure on $T \mathcal{M}$;
\begin{align}
\mathcal{J}_{J_a} = 
\left(
\begin{array}{cc}
J_a & 0 \\
0 & - J^*_a
\end{array}
\right),
\qquad
\mathcal{J}_{\omega_a}
=
\left(
\begin{array}{cc}
0 & - \omega_a^{-1} \\
\omega_a & 0
\end{array}
\right).
\end{align}
For later convenience, we denote 
$\mathcal{J}_{J_a} = \mathcal{J}_{a,+}$ and $\mathcal{J}_{\omega_a} = \mathcal{J}_{a,-}$.
These structures satisfy the algebra
\begin{align}
\mathcal{J}_{a,\pm} \mathcal{J}_{b,\pm} 
= - \delta_{ab} \mathbf{1}_{2D} +
 \epsilon_{abc} \mathcal{J}_{c,+},
\qquad 
\mathcal{J}_{a,\pm} \mathcal{J}_{b,\mp} 
= \delta_{ab} \mathcal{J} +
 \epsilon_{abc} \mathcal{J}_{c,-}.
\label{eq:generalized_hyperkahler_algebra}
\end{align}
Here $\epsilon_{abc}$ is the Levi-Civita symbol and 
$\mathcal{J}$ is the chiral structure 
\begin{align}
\mathcal{J} = 
\left(
\begin{array}{cc}
0 & g^{-1} \\
g & 0
\end{array}
\right),
\end{align}
satisfying $\mathcal{J}^2 = \mathbf{1}_{2D}$.
In fact, the algebra \eqref{eq:generalized_hyperkahler_algebra} is the definition
of the generalized hyperk\"{a}hler structure \cite{Bredthauer:2006sz} 
 and it is the algebra of the split-bi-quaternions or 
$Cl_{0,3} (\mathbb{R})$ in disguise \cite{Kimura:2022dma}.
An algebra that incorporates the split-bi-quaternions and the algebra of
the Born structure (split-quaternions) is 
split-tetra-quaternions.
This is a hypercomplex number generating a 16-dimensional algebra and
isomorphic to 
$Cl_{4,0} (\mathbb{R})$,
$Cl_{1,3} (\mathbb{R})$ and $Cl_{0,4} (\mathbb{R})$.
The split-tetra-quaternions contain the bases of 10 imaginary and 6 real units.
They are represented by (see Appendix \ref{sec:appendix1} and \ref{sec:appendix2})
\begin{alignat}{5}
&
1 e_0 \mathbf{e}_0,
&\qquad
&1 e_0 \mathbf{e}_a,
&\qquad
&1 e_1 \mathbf{e}_0,
&\qquad
&1 e_1 \mathbf{e}_a,
\notag \\
&
i e_2 \mathbf{e}_2,
&\qquad 
&i e_2 \mathbf{e}_a,
&\qquad
&i e_3 \mathbf{e}_0,
&\qquad
&i e_3 \mathbf{e}_a,
&\qquad 
&(a=1,2,3),
\end{alignat}
where $e_{\mu}$ and $\mathbf{e}_{\mu}$ are two commuting quaternions and
$(1,i)$ is the basis of $\mathbb{C}$. Hereafter we omit the ``$1$'' in the products.

We find that the triples $(\mathcal{J}_{a,+}, \mathcal{J}_{a,-}, \mathcal{J})$ $(a=1,2,3)$ 
form three independent bi-complex numbers sharing the common
real structure $\mathcal{J} = \mathcal{J}_{a,+} \mathcal{J}_{a,-} \,
(a:\text{no sum})$.
Since the bi-complex numbers are realized as subalgebras of split-tetra-quaternions as 
\begin{align}
(i e_2 \mathbf{e}_0, e_0 \mathbf{e}_a, i e_2 \mathbf{e}_a),
\qquad 
(i e_3 \mathbf{e}_0, e_0 \mathbf{e}_a, i e_3 \mathbf{e}_a),
\qquad 
(a=1,2,3),
\end{align}
we make an assignment\footnote{
The set ($\mathcal{J} = i e_3 \mathbf{e}_0$, 
$\mathcal{J}_{a,+} = e_0 \mathbf{e}_a$,
$\mathcal{J}_{a,-} = - i e_3 \mathbf{e}_a$)
is an alternative assignment.
};
\begin{align}
\mathcal{J} = i e_2 \mathbf{e}_0,
\qquad 
\mathcal{J}_{a,+} = e_0 \mathbf{e}_a, 
\qquad 
\mathcal{J}_{a,-} = - i e_2 \mathbf{e}_a,
\qquad (a=1,2,3).
\end{align}
The Born structure $(\mathcal{I}, \mathcal{J}, \mathcal{K})$ of
$\mathcal{M}$ obeys the algebra of split-quaternions.
Since the split-quaternion that contains $\mathcal{J}$ is represented by the basis
\begin{align}
(e_1 \mathbf{e}_0, i e_2 \mathbf{e}_0, i e_3 \mathbf{e}_0),
\end{align}
we find the assignment
\begin{align}
\mathcal{I} = e_1 \mathbf{e}_0, 
\qquad 
\mathcal{J} = i e_2 \mathbf{e}_0, 
\qquad 
\mathcal{K} = - i e_3 \mathbf{e}_0.
\end{align}
The other three imaginary units are given by
\begin{align}
ie_3 \mathbf{e}_a, \quad (a=1,2,3)
\end{align}
which are decomposed like $i e_3 \mathbf{e}_a = (e_1 \mathbf{1}) (i e_2 \mathbf{e}_a)$.
Then by assigning the remaining three complex structures 
$\mathcal{P}''$, $\mathcal{Q}''$ and $\mathcal{R}''$ as
\begin{align}
\mathcal{P}'' 
= i e_3 \mathbf{e}_1, \qquad 
\mathcal{Q}'' 
= i e_3 \mathbf{e}_2, \qquad
\mathcal{R}'' 
= i e_3 \mathbf{e}_3,
\end{align}
we find
\begin{align}
\mathcal{P}'' 
= - \mathcal{I} \mathcal{J}_{1,-},
\qquad
\mathcal{Q}'' 
= - \mathcal{I} \mathcal{J}_{2,-},
\qquad
\mathcal{R}'' 
= - \mathcal{I} \mathcal{J}_{3,-}.
\end{align}
Indeed, the direct calculations reveal that they are expressed by
\begin{align}
\mathcal{P}'' =
\left(
\begin{array}{cc}
J_1 & 0 \\
0 & J_1^*
\end{array}
\right),
\qquad 
\mathcal{Q}'' = 
\left(
\begin{array}{cc}
J_2 & 0  \\
0 & J_2^*
\end{array}
\right),
\qquad
\mathcal{R}'' = 
\left(
\begin{array}{cc}
J_3 & 0 \\
0 & J_3^*
\end{array}
\right),
\end{align}
satisfying the desired properties 
$\mathcal{P}''{}^2 = \mathcal{Q}''{}^2 = \mathcal{R}''{}^2 = -\mathbf{1}_{2D}$.
Similarly, we find that the remaining real structures
$\mathcal{S}'', \mathcal{T}'', \mathcal{U}'', \mathcal{V}''$ are given by
\begin{align}
\mathcal{S}'' = e_1 \mathbf{e}_1 = \mathcal{I} \mathcal{J}_{1,+}, 
\qquad 
\mathcal{T}'' = e_1 \mathbf{e}_2 = \mathcal{I} \mathcal{J}_{2,+},
\qquad 
\mathcal{U}'' = e_1 \mathbf{e}_3 = \mathcal{I} \mathcal{J}_{3,+}, 
\qquad
\mathcal{V}'' = e_0 \mathbf{e}_0 = \mathbf{1}_{2D}.
\end{align}

We finally consider the embedding of 
the bi-hypercomplex structure $(J_{a,\pm}, \omega_{a,\pm})$
on $M$ into the doubled structures of $\mathcal{M}$.
The bi-hypercomplex structure
$(I_{a,\pm}, \omega_{a,\pm})$ is embedded into the
generalized hyperk\"{a}hler structure 
\begin{align}
\mathcal{J}_{a,\pm} =
\frac{1}{2}
\Big(
\mathcal{J}_{J_{a,+}} \pm \mathcal{J}_{J_{a,-}} + \mathcal{J}_{\omega_{a,+}}
 \mp \mathcal{J}_{\omega_{a,-}}
\Big).
\end{align}
Note that the K\"{a}hler case corresponds to $J_{a,+} = J_{a,-} = J_a$ and hence
$\mathcal{J}_{a,+} = \mathcal{J}_{J_a}$, $\mathcal{J}_{a,-} = \mathcal{J}_{\omega_a}$.
The generalized hyperk\"{a}hler structure $(\mathcal{J}_{a,+}, \mathcal{J}_{a,-})$ 
forms the algebra of the split-bi-quaternions.
This contains 6 imaginary units $(\mathcal{J}_{a,\pm})^2 = - \mathbf{1}_{2D}$
($a$ :\ no sum).
Furthermore, this contains 6 generalized K\"{a}hler structures
$(\mathcal{J}_{J_{a,\pm}}, \mathcal{J}_{\omega_{a,\pm}}, \mathcal{J})$ 
that share the common real structure $\mathcal{J}$.
Each forms the algebra of the bi-complex numbers.
Then we need 6 bi-complex subalgebras to incorporate these structures.
This is not possible by the split-tetra-quaternions and we therefore enlarge this algebra.
An appropriate algebra is the split-tetra-quaternions over $\mathbb{H}$
(see Appendix \ref{sec:appendix1}).
This is a 64-dimensional algebra on $\mathcal{M}$.
Indeed, if we assign the bases
\begin{align}
\mathcal{J} = i e_2 \mathbf{e}_0 \hat{\mathbf{e}}_0,
\quad
\mathcal{J}_{J_{a,+}} = e_0 \mathbf{e}_a \hat{\mathbf{e}}_0,
\quad
\mathcal{J}_{\omega_{a,+}} = - i e_2 \mathbf{e}_a \hat{\mathbf{e}}_0,
\quad
\mathcal{J}_{J_{a,-}} = e_0 \mathbf{e}_0 \hat{\mathbf{e}}_a,
\quad
\mathcal{J}_{\omega_{a,-}} = - i e_2 \mathbf{e}_0 \hat{\mathbf{e}}_a,
\end{align}
then we find that the triples 
\begin{align}
\left(
\mathcal{J}_{J_{a,+}}, \mathcal{J}_{\omega_{a,+}}, \mathcal{J}
\right),
\quad
\left(
\mathcal{J}_{J_{a,-}}, \mathcal{J}_{\omega_{a,-}}, \mathcal{J}
\right)
\end{align}
obey the algebra of bi-complex numbers.
In this basis, $\mathcal{J}_{a,\pm}$ are represented by
\begin{align}
\mathcal{J}_{a,+}
=& \ 
\frac{1}{2}
\Big[
e_0 (\mathbf{e}_a \hat{\mathbf{e}}_0 + \mathbf{e}_0 \hat{\mathbf{e}}_a)
- i e_2 
(\mathbf{e}_a \hat{\mathbf{e}}_0 - \mathbf{e}_0 \hat{\mathbf{e}}_a)
\Big],
\notag \\
\mathcal{J}_{a,-}
=& \ 
\frac{1}{2}
\Big[
e_0 (\mathbf{e}_a \hat{\mathbf{e}}_0 - \mathbf{e}_0 \hat{\mathbf{e}}_a)
- i e_2 
(\mathbf{e}_a \hat{\mathbf{e}}_0 + \mathbf{e}_0 \hat{\mathbf{e}}_a)
\Big].
\end{align}
Using these expressions, we compute
\begin{align}
\mathcal{J}_{a,+} \mathcal{J}_{b,+}  
=& \ 
- \delta_{ab} e_0 \mathbf{e}_0 \hat{\mathbf{e}}_0
+ \epsilon_{abc}
 \frac{1}{2}
\Big[
e_0 (\mathbf{e}_c \hat{\mathbf{e}}_0 + \mathbf{e}_0 \hat{\mathbf{e}}_c)
- i e_2 (\mathbf{e}_c \hat{\mathbf{e}}_0 - \mathbf{e}_0 \hat{\mathbf{e}}_c)
\Big]
\notag \\
=& \ - \delta_{ab} \mathbf{1}_{2D} + \epsilon_{abc} \mathcal{J}_{c,+},
\notag \\
\mathcal{J}_{a,-} \mathcal{J}_{b,-} 
=& \ 
- \delta_{ab} e_0 \mathbf{e}_0 \hat{\mathbf{e}}_0
+ \epsilon_{abc}
 \frac{1}{2}
\Big[
e_0 (\mathbf{e}_c \hat{\mathbf{e}}_0 + \mathbf{e}_0 \hat{\mathbf{e}}_c)
- i e_2 (\mathbf{e}_c \hat{\mathbf{e}}_0 - \mathbf{e}_0 \hat{\mathbf{e}}_c)
\Big]
\notag \\
=& \ - \delta_{ab} \mathbf{1}_{2D} + \epsilon_{abc} \mathcal{J}_{c,+},
\notag \\
\mathcal{J}_{a,\pm} \mathcal{J}_{b,\mp} 
=& \ 
\delta_{ab} i e_2 \mathbf{e}_0 \hat{\mathbf{e}}_0 + \epsilon_{abc}
\frac{1}{2}
\Big[
e_0 (\mathbf{e}_c \hat{\mathbf{e}}_0 - \mathbf{e}_0 \hat{\mathbf{e}}_c)
- i e_2
(\mathbf{e}_c \hat{\mathbf{e}}_0 + \mathbf{e}_0 \hat{\mathbf{e}}_c)
\Big]
\notag \\
=& \ \delta_{ab} \mathcal{J} + \epsilon_{abc} \mathcal{J}_{c,-}.
\end{align}
Here we have used the fact that 
$e_{\mu}, \mathbf{e}_{\mu}, \hat{\mathbf{e}}_{\mu}$ 
are commuting quaternions and denoted 
$e_0 \mathbf{e}_0 \hat{\mathbf{e}}_0 = \mathbf{1}_{2D}$.
Then we confirm the algebra of the split-bi-quaternions for the
generalized hyperk\"{a}hler structure.

In summary, the hyperk\"{a}hler structure 
$(J_a, \omega_a)$ on $M$ is embedded into
 the doubled structure on $\mathcal{M}$ that obey the 16-dimensional
 algebra of the  split-tetra-quaternions.
The bi-hypercomplex structure 
$(J_{a,\pm}, \omega_{a,\pm})$ is embedded into the doubled
 structure that obeys the 64-dimensional algebra of the split-tetra-quaternions over $\mathbb{H}$.
The algebras of doubled structures on $T\mathcal{M} \simeq TM \oplus T^*M$ 
are summarized in Table \ref{tb:generalized_structure_algebra}.
\begin{table}[t]
\centering
\scalebox{.905}{%
\begin{tabular}{l|l|l}
Structures on $T\mathcal{M} \simeq TM \oplus T^*M$ & Algebras of hypercomplex numbers & Structures on $TM$ \\
\hline
\hline
Generalized K\"{a}hler & bi-complex numbers (4) & K\"ahler $(J, \omega)$ \\ 
\hline
Generalized K\"{a}hler & bi-complex numbers over $\mathbb{C}$ (8) & bi-hermitian $(J_{\pm}, \omega_{\pm})$ \\ 
\hline
Generalized hyperk\"{a}hler  & split-bi-quaternions (8) & hyperk\"{a}hler $(J_a, \omega_a)$ \\ 
\hline
Generalized hyperk\"{a}hler  & split-bi-quaternions over $\mathbb{H}$ (32) & bi-hypercomplex $(J_{a,\pm}, \omega_{a,\pm})$ \\ 
\hline
Born & split-quaternions (4) &  \\ 
\hline
Born + generalized K\"{a}hler & bi-quaternions (8) & K\"{a}hler $(J, \omega)$ \\
\hline
Born + generalized K\"{a}hler & bi-quaternions over $\mathbb{C}$ (16) & bi-hermitian $(J_{\pm}, \omega_{\pm})$ \\ 
\hline
Born + generalized hyperk\"{a}hler & split-tetra-quaternions (16) & hyperk\"{a}hler $(J_a, \omega_a)$ \\ 
\hline
Born + generalized hyperk\"{a}hler & split-tetra-quaternions over $\mathbb{H}$ (64) & bi-hypercomplex $(J_{a,\pm}, \omega_{a,\pm})$ \\ 
\end{tabular}
}
\caption{The structures on $T\mathcal{M} \simeq TM \oplus T^*M$ and their algebras and dimensions.}
\label{tb:generalized_structure_algebra}
\end{table}

\section{Worldsheet instantons in Born sigma models} \label{sec:Born_sigma_model}
We have established the T-duality covariant embeddings of complex
structures of spacetime.
One of the notions that has deep connections with the spacetime complex
structures is the worldsheet instantons \cite{Wen:1985jz, Dine:1986zy}.
In \cite{Kimura:2022dma}, we 
studied the T-duality relation between the
instantons in K\"{a}hler and bi-hermitian geometries.
Due to the fact that there are complex structures $J$ and $J_{\pm}$ in the K\"{a}hler
and the bi-hermitian geometries respectively, 
we find a one-to-two correspondence between the instantons in each geometry.
We here elucidate this relation within the T-duality covariant formulation.

In this section, we study the worldsheet instantons in a T-duality
covariant doubled formalism.
The doubled formalism of string sigma models that makes T-duality be manifest
has been studied in various viewpoints 
\cite{Tseytlin:1990nb, Hull:2004in,Hull:2006va, Copland:2011wx}.
Among other things, more direct connections to DFT and the doubled space appear in the Born
sigma model \cite{Marotta:2019eqc}\footnote{
See \cite{Sakatani:2020umt, Marotta:2022tfe} for generalizations to
branes and exceptional geometries.
}.
In the following, we show that the worldsheet instanton equations respecting T-duality
symmetry are obtained in the Born sigma model by utilizing the doubled
structures discussed in the previous sections.

\subsection{Born sigma models}
The Born sigma model is a sigma model whose target space is the Born manifold $\mathcal{M}$.
This is closely related to the doubled sigma model of a string introduced in \cite{Hull:2004in,Hull:2006va}.
The action of the Born sigma model in the Minkowski signature is given by 
\begin{align}
S = \frac{1}{4} \int_{\Sigma} \,
\left[
\mathcal{H}_{MN} {\dop} \mathbb{X}^M \wedge * {\dop} \mathbb{X}^N
-
\Omega_{MN} {\dop} \mathbb{X}^M \wedge {\dop} \mathbb{X}^N
\right].
\label{eq:Born_sigma_model}
\end{align}
Here $\Sigma$ is the two-dimensional worldsheet, 
$\mathbb{X}^M = (X^{\mu},\tilde{X}_{\mu})$ is the local coordinate of
the Born manifold $\mathcal{M}$, $*$ is the Hodge star operator in
$\Sigma$, $\mathcal{H}_{MN}$ is the generalized metric in the Born manifold
$\mathcal{M}$ and $\Omega_{MN} = - \Omega_{NM}$ is an anti-symmetric
constant matrix. We have neglected the Fradkin-Tseytlin term that
involves the dilaton \cite{Marotta:2019eqc} which is not relevant in our
discussion.
The action \eqref{eq:Born_sigma_model} is invariant under the $O(D,D)$ rotation
\begin{gather}
{\dop} \mathbb{X}^M \to \mathcal{O}^M{}_N {\dop} \mathbb{X}^N,
\qquad
\mathcal{H}_{MN} \to (\mathcal{O}^t)_M{}^P \mathcal{H}_{PQ} \mathcal{O}^Q{}_N,
\qquad 
\Omega_{MN} \to (\mathcal{O}^t)_M{}^P \Omega_{PQ} \mathcal{O}^Q{}_N,
\notag \\
\mathcal{O} \in O(D,D).
\end{gather}
In the following, we use the standard parameterization of the generalized
metric \eqref{eq:DFT_parametrization}.
The second term in the action \eqref{eq:Born_sigma_model} 
is topological but plays an important role
in the instanton equations.
Following \cite{Hull:2004in,Hull:2006va}, we employ the expression
of the topological term 
$\Omega_{MN} {\dop} \mathbb{X}^M \wedge {\dop} \mathbb{X}^N = - 2 {\dop} X^{\mu} \wedge {\dop} \tilde{X}_{\mu}$.

Using the standard parameterization, 
the term involving $\mathcal{H}_{MN}$ in the action \eqref{eq:Born_sigma_model} is expanded as
\begin{align}
\frac{1}{4} \int_{\Sigma} \! \mathcal{H}_{MN} \dop \mathbb{X}^M \wedge *
 \dop \mathbb{X}^N
=& \ \frac{1}{4} \int_{\Sigma} \! {\dop}^2 \sigma \, \sqrt{-h} h^{\alpha \beta}
\Big[
\left(
g_{\mu \nu} - B_{\mu \rho} g^{\rho \sigma} B_{\sigma \nu}
\right) \del_{\alpha} X^{\mu} \del_{\beta} X^{\nu}
- g^{\mu \rho} B_{\rho \nu} \del_{\alpha} \tilde{X}_{\mu} \del_{\beta} X^{\nu}
\notag \\
& \qquad \qquad \qquad \qquad 
+ B_{\mu \rho} g^{\rho \nu} \del_{\alpha} X^{\mu} \del_{\beta} \tilde{X}_{\nu}
+ g^{\mu \nu} \del_{\alpha} \tilde{X}_{\mu} \del_{\beta} \tilde{X}_{\nu}
\Big],
\label{eq:doubled_sigma_model}
\end{align}
where $h_{\alpha \beta}, \, (\alpha, \beta = 0,1)$ is the metric of the two-dimensional worldsheet
$\Sigma$ and the topological term is
\begin{align}
- \frac{1}{4} \int_{\Sigma} {\dop}^2 \sigma \, \sqrt{-h} \, 
\varepsilon^{\alpha \beta}
 \Omega_{MN} \del_{\alpha} \mathbb{X}^M \del_{\beta} \mathbb{X}^N
= 
 \frac{1}{2} \int_{\Sigma} \! {\dop}^2 \sigma \, 
\epsilon^{\alpha \beta}
 \Big[
 \del_{\alpha} X^{\mu} \del_{\beta} \tilde{X}_{\mu}
 \Big].
\end{align}
Here 
$\varepsilon^{\alpha \beta}$
and 
$\epsilon^{\alpha \beta}$
are the totally anti-symmetric tensor and
the Levi-Civita symbol in $\Sigma$, respectively.

Since the Born sigma model \eqref{eq:Born_sigma_model} contains double
degrees of freedom, we impose constraints on the quantities.
The physical (non-doubled) sigma model is obtained by imposing the DFT
constraints \eqref{eq:DFT_constraints} on the background fields 
$g_{\mu \nu}$, $B_{\mu \nu}$, $\phi$ and also 
by introducing the self-duality condition;
\begin{align}
{\dop} \mathbb{X}^M =  \eta^{MP} \mathcal{H}_{PQ} * {\dop} \mathbb{X}^Q.
\end{align}
This is rewritten by the chiral structure $\mathcal{J} = \eta^{-1} \mathcal{H}$ in $\mathcal{M}$ as
\begin{align}
{\dop} \mathbb{X}^M = \mathcal{J}^M {}_N * {\dop} \mathbb{X}^N.
\label{eq:chirality}
\end{align}
Therefore \eqref{eq:chirality} is just the chirality condition.
By using the representation \eqref{eq:Born_structure_representation2} for
the chiral structure $\mathcal{J}$, the condition \eqref{eq:chirality}
is expanded and ${\dop} \tilde{X}_{\mu}$ is solved as
\begin{align}
{\dop} \tilde{X}_{\mu} = g_{\mu \nu} * {\dop} X^{\nu} + B_{\mu \nu} {\dop} X^{\nu}.
\label{eq:chirality_solution}
\end{align}
Then we can remove the winding degrees of freedom ${\dop} \tilde{X}_{\mu}$
from the action.
Note that we have solved the DFT constrains \eqref{eq:DFT_constraints}
by making all the background fields depend only on $X^{\mu}$.
Plugging \eqref{eq:chirality_solution} back into the action
\eqref{eq:Born_sigma_model}, we find the $\mathcal{H}_{MN}$ part vanishes.
On the other hand, the topological term becomes
\begin{align}
- \frac{1}{4} \Omega_{MN} {\dop} \mathbb{X}^M \wedge {\dop} \mathbb{X}^N
=& \  
\frac{1}{2}
{\dop} X^{\mu} \wedge 
\Big[
g_{\mu \rho} * {\dop} X^{\rho} + B_{\mu \rho} {\dop} X^{\rho}
\Big]
\notag \\
=& \ \frac{1}{2} g_{\mu \nu} {\dop} X^{\mu} \wedge * {\dop} X^{\nu}
+ \frac{1}{2} B_{\mu \nu} {\dop} X^{\mu} \wedge {\dop} X^{\nu}.
\end{align}
This precisely reproduces the action of the ordinary string sigma model.

\subsection{Instantons in Born sigma models}
We then consider the instantons in the Born sigma model.
In the following, spacetime and the worldsheet have the Euclidean
signature and $*^2 = -1$.
We first consider the term that depends on $\mathcal{H}_{MN}$ in the action
\eqref{eq:Born_sigma_model}.
Since the metric $\mathcal{H}_{MN}$ is positive-definite 
in the Euclidean space, we have the Bogomol'nyi bound of the action;
\begin{align}
S_{\text{E}} =& \ \frac{1}{8} \int_{\Sigma} \! {\dop}^2 \sigma \,
\sqrt{h}
\Big[
h^{\alpha \beta} \mathcal{H}_{MN} 
\left(
\del_{\alpha} \mathbb{X}^M \pm \mathcal{A}^M {}_P 
\varepsilon_{\alpha \gamma}
\del^{\gamma} \mathbb{X}^P
\right)
\left(
\del_{\beta} \mathbb{X}^N \pm \mathcal{A}^N {}_Q 
\varepsilon_{\beta \delta}
\del^{\delta} \mathbb{X}^Q
\right)
\notag \\
& \qquad \qquad \qquad \quad 
\pm
2 (\omega_{\mathcal{A}} )_{MN} 
\varepsilon^{\alpha \beta}
\del_{\alpha} \mathbb{X}^M \del_{\beta} \mathbb{X}^N
\Big]
\notag \\
\ge& \ 
\pm
\frac{1}{4} \int_{\Sigma} \! {\dop}^2 \sigma \,
\sqrt{h}
 (\omega_{\mathcal{A}})_{MN} 
\varepsilon^{\alpha \beta}
\del_{\alpha} \mathbb{X}^M \del_{\beta}
 \mathbb{X}^N,
\end{align}
where $\mathcal{A}$ is a doubled structure satisfying 
$\mathcal{A}^2 = - \mathbf{1}_{2D}$ in the Born manifold $\mathcal{M}$ and
$\omega_{\mathcal{A}} = \mathcal{H} \mathcal{A}$ is the fundamental
two-form associated with $\mathcal{A}$.
The bound is saturated when the map $\mathbb{X} : \Sigma \to \mathcal{M}$ satisfies
\begin{align}
\del_{\alpha} \mathbb{X}^M \pm \mathcal{A}^M {}_N 
\varepsilon_{\alpha \beta}
\del^{\beta} \mathbb{X}^N = 0,
\label{eq:instanton}
\end{align}
or equivalently,
\begin{align}
{\dop} \mathbb{X}^M \pm \mathcal{A}^M {}_N * {\dop} \mathbb{X}^N = 0.
\label{eq:doubled_instanton_eq}
\end{align}
We call these the doubled instanton equations.
By the chirality condition \eqref{eq:chirality}\footnote{
Note that we should replace $* \to - i *$ in \eqref{eq:chirality} for the Euclidean space.
}
and the doubled instanton equations, we have
\begin{align}
{\dop} \mathbb{X}^M = \mp \mathcal{A}^M{}_N * {\dop} \mathbb{X}^N
= \mp i \mathcal{A}^M{}_N \mathcal{J}^N{}_P {\dop} \mathbb{X}^P
= ((\mp i \mathcal{A} \mathcal{J})^2){}^M{}_N {\dop} \mathbb{X}^N.
\end{align}
This means that we need $(\mathcal{A} \mathcal{J})^2 = - \mathbf{1}_{2D}$ to have non-trivial
solutions for instantons, otherwise $\mathbb{X}^I = 0$.
Since $\mathcal{A}^2 = - \mathbf{1}_{2D}$ and $\mathcal{J}^2 = \mathbf{1}_{2D}$, 
we obtain $(\mathcal{A} \mathcal{J})^2 = - \mathbf{1}_{2D}$
if ${\mathcal{A}}$ commutes with $\mathcal{J}$ (i.e., $[\mathcal{A}, \mathcal{J}] = 0$).
On the other hand, if $\mathcal{A}$ anti-commutes with $\mathcal{J}$
(i.e., $\{ \mathcal{A}, \mathcal{J}\} = 0$), then we find $(\mathcal{A} \mathcal{J})^2 = \mathbf{1}_{2D}$.
Therefore only $\mathcal{A}$ that commutes with the chiral structure
$\mathcal{J}$ is allowed for the doubled instantons.

In the following, we consider $\Sigma = S^2$ and the image of the map 
$\mathbb{X}$ is a two-cycle $\mathcal{C}^2$ in $\mathcal{M}$.
Then the map $\mathbb{X}: \Sigma \to \mathcal{M}$ is classified by the
homotopy class $\pi_2 (S^2)$.
The action bound is given by 
\begin{align}
S_{\text{E}} = 
\left|
\frac{1}{4} \int_{\Sigma} (\omega_{\mathcal{A}})_{MN} {\dop} \mathbb{X}^M \wedge {\dop} \mathbb{X}^N
\right|
+ \frac{i}{4} \int_{\Sigma} \Omega_{MN} {\dop} \mathbb{X}^M \wedge {\dop} \mathbb{X}^N
=
\frac{1}{4}
\left|
\int_{\mathcal{C}^2} \! \omega_{\mathcal{A}}
\right|
+ \frac{i}{4} \int_{\mathcal{C}^2} \Omega.
\label{eq:general_action_bound}
\end{align}
Here we have restored the topological term $\int \Omega$.
We assume that the two-cycle $\mathcal{C}^2$ lie 
in the physical spacetime $M$.

Note that the topological term is written in the T-duality covariant form;
\begin{align}
\Omega_{MN} {\dop} \mathbb{X}^M \wedge {\dop} \mathbb{X}^N
=& \ 
\Omega_{MN} 
\left(
\mp \mathcal{A}^M {}_P * {\dop} \mathbb{X}^P
\right) \wedge {\dop} \mathbb{X}^N
\notag \\
=& \ 
\mp i \Omega_{MP} \mathcal{A}^P {}_Q \mathcal{J}^Q {}_N \,
{\dop} \mathbb{X}^M \wedge {\dop} \mathbb{X}^N.
\end{align}

In the following, we first study the doubled instanton equations for the
K\"{a}hler geometry and then move to the bi-hermitian geometry.
For simplicity we start 
by the K\"{a}hler geometry with trivial
$B$-field $B = 0$.
There 
is a doubled structure
$(\mathcal{J}_J, \mathcal{J}_{\omega}, \mathcal{I}, \mathcal{Q})$
in the doubled space whose squares are $-\mathbf{1}_{2D}$.
As we have shown, $\mathcal{I}$ and $\mathcal{Q}$ anti-commute with the
chiral structure $\mathcal{J}$ and they give trivial solutions to the
equations \eqref{eq:doubled_instanton_eq}.
On the other hand we have 
$[\mathcal{J}_{J}, \mathcal{J}] = [\mathcal{J}_{\omega}, \mathcal{J}] = 0$
and the doubled instantons
defined by $\mathcal{J}_J, \mathcal{J}_{\omega}$ provide meaningful solutions.
We clarify them explicitly.

\paragraph{$\mathcal{A} = \mathcal{J}_{\omega}$ case.}
In the case of $\mathcal{A} = \mathcal{J}_{\omega}$, the doubled
instanton equations are
\begin{align}
{\dop} \mathbb{X}^M \pm (\mathcal{J}_{\omega})^M {}_N * {\dop} \mathbb{X}^N = 0,
\label{eq:doubled_instantons_omega}
\end{align}
which in components are written as 
\begin{align}
&
\del_{\alpha} X^{\mu} \pm (- (\omega^{-1})^{\mu \nu}) 
\varepsilon_{\alpha \beta}
 \del^{\beta} \tilde{X}_{\nu} = 0,
\notag \\
&
\del_{\alpha} \tilde{X}_{\mu} \mp \omega_{\mu \nu} 
\varepsilon_{\alpha \beta}
 \del^{\beta} X^{\nu} = 0.
\label{eq:doubled_instanton_Jomega}
\end{align}
Here $\omega = - g J$ is the K\"{a}hler form in $M$.
By the chirality condition \eqref{eq:chirality}, we solve the winding
coordinate as 
$\del_{\alpha} \tilde{X}^{\mu} = - i 
\varepsilon_{\alpha \beta}
g_{\mu \nu} \del^{\beta} X^{\nu}$.
Plugging this into the first line in
\eqref{eq:doubled_instanton_Jomega}, we obtain
\begin{align}
J^{\mu} {}_{\nu} \del_{\alpha} X^{\nu} = \mp i \del_{\alpha} X^{\mu}.
\label{eq:holomorphic_vectors}
\end{align}
In general, 
the almost complex structure $J$ of the spacetime manifold $M$
decomposes the complexified tangent space 
$TM^{\mathbb{C}} = TM \otimes \mathbb{C}$ into $TM^{\pm}$ such as 
\begin{align}
TM^{\mathbb{C}} = TM^+ \oplus TM^-.
\end{align}
Here $TM^{\pm}$ are eigenbundles of the complex structure 
$J X_{\pm} = \pm i X_{\pm}$ and 
$X_+$, $X_-$ are (anti)holomorphic vectors.
When $J$ is integrable, the Lie bracket of the (anti)holomorphic vectors
become the (anti)holomorphic vectors.
Therefore the doubled instanton equations by $\mathcal{J}_{\omega}$ restrict
${\dop} X^{\mu}$ to (anti)holomorphic vectors by $J$.

The fundamental two-form $\omega_{\mathcal{J}_{\omega}}$ becomes
\begin{align}
\omega_{\mathcal{J}_{\omega}} = \mathcal{H} \mathcal{J}_{\omega}
= 
\left(
\begin{array}{cc}
g & 0  \\
0 & g^{-1}
\end{array}
\right)
\left(
\begin{array}{cc}
0 & - \omega^{-1} \\
\omega & 0
\end{array}
\right)
=
\left(
\begin{array}{cc}
0 & - g \omega^{-1} \\
g^{-1} \omega & 0 
\end{array}
\right).
\end{align}
Then the term 
$|\int_{C^2} \omega_{\mathcal{J}_{\omega}}|$
in
the action bound \eqref{eq:general_action_bound}
becomes trivial;
\begin{align}
(\omega_{\mathcal{J}_{\omega}})_{MN} {\dop} \mathbb{X}^M \wedge {\dop} \mathbb{X}^N
= (g_{\mu \rho} \omega^{\rho \nu} + \omega_{\mu \rho} g^{\nu \rho}) 
{\dop} X^{\mu} \wedge {\dop} \tilde{X}_{\nu}
= 0.
\end{align}
Here we have used the relation 
$g \omega^{-1} + \omega g^{-1} = - g J^{-1} g^{-1} - g J g^{-1} = 0$.
The non-trivial action
bound comes from the topological term $\Omega$
and it is nothing but the (Euclideanized) string sigma model action;
\begin{align}
S_{\text{E}} = \frac{1}{2} \int \! g_{\mu \nu} {\dop} X^{\mu} \wedge * {\dop} X^{\nu} 
+ \frac{i}{2} \int \! B_{\mu \nu} {\dop} X^{\mu} \wedge {\dop} X^{\nu}.
\end{align}

\paragraph{$\mathcal{A} = \mathcal{J}_J$ case.}
When $\mathcal{A} = \mathcal{J}_J$, the doubled instanton equations become
\begin{align}
{\dop} \mathbb{X}^M \pm (\mathcal{J}_J)^M {}_N * {\dop} \mathbb{X}^N = 0.
\label{eq:doubled_instantons_kahler}
\end{align}
In components, these are given by
\begin{align}
\del_{\alpha} X^{\mu} \pm J^{\mu} {}_{\nu} 
\varepsilon_{\alpha \beta}
 \del^{\beta} X^{\nu} &= 0,
\notag \\
\del_{\alpha} \tilde{X}_{\mu} \mp J^*_{\mu} {}^{\nu} 
\varepsilon_{\alpha \beta}
 \del^{\beta} \tilde{X}_{\nu} &= 0.
\label{eq:doubled_instanton_decomp}
\end{align}
The first equation reproduces the ordinary worldsheet instanton
equations \cite{Wen:1985jz}.
The second equation provides us the T-dual of the first equation.
We find that the chirality condition 
$\del_{\alpha} \tilde{X}_{\mu} = - i
\varepsilon_{\alpha \beta}
g_{\mu \nu} \del^{\beta} X^{\nu}$ 
applying to
the second equation gives the first one.
In this case, the fundamental two-form $\omega_{\mathcal{J}_J}$ is
evaluated as 
\begin{align}
\omega_{\mathcal{J}_J} = \mathcal{H} \mathcal{J}_J = 
\left(
\begin{array}{cc}
g J & 0  \\
0 & - g^{-1} J^*
\end{array}
\right) 
= 
\left(
\begin{array}{cc}
- \omega & 0  \\
0 &  \omega^{-1}
\end{array}
\right).
\end{align}
Therefore we have
\begin{align}
\frac{1}{4} 
\int_{\mathcal{C}^2} \omega_{\mathcal{J}_I} = 
\frac{1}{4} 
\int_{\mathcal{C}^2}
\Big[
 \omega_{\mu \nu} {\dop} X^{\mu} \wedge {\dop} X^{\nu}
- (\omega^{-1})^{\mu \nu} {\dop} \tilde{X}_{\mu} \wedge {\dop} \tilde{X}_{\nu}
\Big].
\end{align}
By using the chirality condition and eliminating $\tilde{X}_{\mu}$
sector, we have
\begin{align}
(\omega^{-1})^{\mu \nu} {\dop} \tilde{X}_{\mu} \wedge {\dop} \tilde{X}_{\nu}
= 
\omega_{\mu \nu} \varepsilon_{\alpha \beta} \del^{\alpha} X^{\mu} \del^{\beta} X^{\nu}.
\end{align}
Here we have used the relations $\omega = - g J$ and
$\omega^{-1} = - J^{-1} g^{-1} =  J g^{-1}$.
Then, we find the action bound coming from the $\mathcal{H}$ part is
\begin{align}
\frac{1}{4} \int_{\mathcal{C}^2} \omega_{\mathcal{J}_J} = 0.
\end{align}
This is anticipated because the chirality condition makes the
$\mathcal{H}_{MN}$ part be trivial.
On the other hand, the topological term in the action is evaluated as 
\begin{align}
\frac{i}{4} \int \Omega =& \ 
- \frac{i}{2} \int {\dop} X^{\mu} \wedge \Big( -i * g_{\mu \nu} {\dop} X^{\nu} \Big)
\notag \\
=& \ - \frac{1}{2} \int {\dop} X^{\mu} \wedge \big( \pm g_{\mu\nu} J^{\nu} {}_{\rho} {\dop} X^{\rho} \big)
\notag \\
=& \ \pm \frac{1}{2} \int_{\mathcal{C}^2} \omega_{\mu \nu} {\dop} X^{\mu} \wedge {\dop} X^{\nu}.
\end{align}
This reproduces the action
bound of the ordinary worldsheet instantons.

For later convenience, we here introduce the $B$-field by the $B$-transformation.
For a doubled structure $\mathcal{A}$, the action bound is found to be
\begin{align}
S_{\text{E}} \ge \pm \frac{1}{4} \int_{\Sigma} \! {\dop}^2 \sigma \, \sqrt{h}
 (\omega_{\mathcal{A}^B})_{MN} 
\varepsilon^{\alpha \beta}
\del_{\alpha} \mathbb{X}^M \del_{\beta} \mathbb{X}^N,
\end{align}
where the doubled structure $\mathcal{A}$ is replaced by 
$\mathcal{A}^B = e^B \mathcal{A} e^{-B}$ and 
$\omega_{\mathcal{A}^B}$ is the fundamental two-form
associated with $\mathcal{A}^B$.
The general doubled instanton equations are then given by
\begin{align}
{\dop} \mathbb{X}^M \pm (\mathcal{A}^B)^M {}_N * {\dop} \mathbb{X}^N = 0.
\end{align}

For the generalized complex structure $\mathcal{A} = \mathcal{J}_{\omega}$, we have
\begin{align}
\mathcal{J}_{\omega}^B = 
\begin{pmatrix}
(\omega^{-1} B)^{\mu} {}_{\nu} & - (\omega^{-1})^{\mu \nu} \\
\omega_{\mu \nu} + (B \omega^{-1} B)_{\mu \nu} & - (B \omega^{-1})_{\mu} {}^{\nu}
\end{pmatrix}
.
\end{align}
The doubled instanton equations in components become
\begin{gather}
{\dop} X^{\mu} \pm 
\left\{
(\omega^{-1} B)^{\mu} {}_{\nu} * {\dop} X^{\nu} - (\omega^{-1})^{\mu \nu} * {\dop} \tilde{X}_{\nu}
\right\} = 0,
\notag \\
{\dop} \tilde{X}_{\mu} \pm 
\left\{
\omega_{\mu} * {\dop} X^{\nu} + (B \omega^{-1} B)_{\mu \nu} * {\dop} X^{\nu} 
- (B \omega^{-1})_{\mu} {}^{\nu} * {\dop} \tilde{X}_{\nu} 
\right\} = 0.
\label{eq:omega_B_inst}
\end{gather}
Using the chiral structure 
\begin{align}
\mathcal{J}^B = 
\begin{pmatrix}
- g^{-1} B & - g^{-1} \\
g + B g^{-1} B & - B g^{-1}.
\end{pmatrix}
,
\end{align}
the chirality condition is solved by
\begin{align}
{\dop} \tilde{X}_{\mu} = - i g_{\mu \nu} * {\dop} X^{\nu} + B_{\mu \nu} {\dop} X^{\nu}.
\end{align}
By substituting this into the first line in \eqref{eq:omega_B_inst},
we find the condition \eqref{eq:holomorphic_vectors} obtained in the
case of the trivial $B$-field.
In the second line, we have 
\begin{align}
- i g_{\mu \nu} * 
\left(
{\dop} X^{\nu} \mp i J^{\nu} {}_{\rho} {\dop} X^{\rho}
\right)
+ B_{\mu \nu} 
\left(
{\dop} X^{\nu} \mp i J^{\nu} {}_{\rho} {\dop} X^{\rho}
\right) = 0.
\end{align}
This again implies the condition \eqref{eq:holomorphic_vectors}.
Therefore ${\dop} X^{\mu}$ is a (anti)holomorphic vector even in the presence
of the $B$-field. 
The action bound is similarly obtained.

For the case $\mathcal{A} = \mathcal{J}_J$, we have
\begin{align}
\mathcal{J}_{J}^B = e^B \mathcal{J}_J e^{-B} = 
\begin{pmatrix}
J & 0  \\
BJ + J^*B & - J^*
\end{pmatrix}.
\end{align}
The doubled instanton equations are then
\begin{align}
{\dop} \mathbb{X}^M \pm (\mathcal{J}^B_{J})^M {}_N * {\dop} \mathbb{X}^N = 0.
\end{align}
In components, we have
\begin{gather}
{\dop} X^{\mu} \pm J^{\mu} {}_{\nu} * {\dop} X^{\nu} = 0, 
\notag \\
{\dop} \tilde{X}_{\mu} \pm (BJ + J^*B)_{\mu \nu} * {\dop} X^{\nu} \mp (J^*)_{\mu}{}^{\nu} * {\dop} \tilde{X}_{\nu} = 0.
\label{eq:d_inst_B2}
\end{gather}
The first line gives the worldsheet instanton equation.
Under the chirality condition, the second line
in \eqref{eq:d_inst_B2} becomes
\begin{align}
0 = & \
\left(
{\dop} X^{\mu} \pm J^{\mu} {}_{\nu} * {\dop} X^{\nu}
\right)
+ (g^{-1} B)^{\mu} {}_{\nu} * 
\left(
{\dop} X^{\nu} \pm J^{\nu} {}_{\rho} * {\dop} X^{\rho}
\right).
\end{align}
Therefore \eqref{eq:d_inst_B2} consistently reproduces the instanton equation 
${\dop} X^{\mu} \pm J^{\mu} {}_{\nu} * {\dop} X^{\nu} = 0$
even in the presence of the $B$-field.
In this case, the action bound is given by
\begin{align}
S_{\text{E}} = \pm \frac{1}{2} \int \! \omega_{\mu \nu} {\dop} X^{\mu} \wedge {\dop} X^{\nu} 
+ \frac{i}{2} \int \! B_{\mu \nu} {\dop} X^{\mu} \wedge {\dop} X^{\nu}.
\end{align}
Then the topological $\theta$-term for the instanton bound is precisely
obtained by the $B$-field.

\paragraph{Bi-hermitian geometry.}
We next consider the bi-hermitian geometry characterized by $(J_+, J_-)$.
It is known that K\"{a}hler and bi-hermitian geometries are T-dual
with each other.
This becomes apparent when these structures are embedded into
generalized K\"{a}hler structures in the doubled space
\cite{Kimura:2022dma}.
Indeed, the doubled instanton equations in the bi-hermitian geometry are
obtained by the T-duality transformation of \eqref{eq:doubled_instantons_kahler}.
The bi-hermitian structure $(J_+,J_-)$ on spacetime $M$ is expressed
by the generalized complex structures as 
\begin{align}
\mathcal{J}^B_{\pm} =& \ 
\frac{1}{2} 
\begin{pmatrix}
1 & 0 \\
B & 1
\end{pmatrix}
\begin{pmatrix}
J_+ \pm J_- & - (\omega^{-1}_+ \mp \omega_-^{-1}) \\
\omega_+ \mp \omega_- & - (J^*_+ \pm J^*_-)
\end{pmatrix}
\begin{pmatrix}
1 & 0 \\
- B & 1
\end{pmatrix}
\notag \\
=& \ 
\frac{1}{2} 
\Big(
\mathcal{J}^B_{J_+} \pm \mathcal{J}^B_{J_-}
+
\mathcal{J}^B_{\omega_+} \mp \mathcal{J}^B_{\omega_-}
\Big).
\end{align}
The doubled instanton equations are given by 
\begin{align}
{\dop} \mathbb{X}^M \pm (\mathcal{J}^B_{+})^M {}_N * {\dop} \mathbb{X}^N &= 0,
\label{eq:d_inst_bh}
\notag \\
{\dop} \mathbb{X}^M \pm (\mathcal{J}^B_{-})^M {}_N * {\dop} \mathbb{X}^N &= 0.
\end{align}
In the following, we focus on $\mathcal{J}^B_+$ without loss of generality.
The equations \eqref{eq:d_inst_bh} are decomposed as
\begin{align}
& 
\frac{1}{2} 
\Big(
{\dop} \mathbb{X}^M  \pm (\mathcal{J}^B_{J_+})^M{}_N * {\dop} \mathbb{X}^N
\Big)
+
\frac{1}{2} 
\Big(
{\dop} \mathbb{X}^M  \pm (\mathcal{J}^B_{J_-})^M{}_N * {\dop} \mathbb{X}^N
\Big)
\notag \\
& \qquad 
+ 
\frac{1}{2} 
\Big(
{\dop} \mathbb{X}^M  \pm (\mathcal{J}^B_{\omega_+})^M{}_N * {\dop} \mathbb{X}^N
\Big)
-
\frac{1}{2} 
\Big(
{\dop} \mathbb{X}^M  \pm (\mathcal{J}^B_{\omega_-})^M{}_N * {\dop} \mathbb{X}^N
\Big) = 0.
\end{align}
Then the equations \eqref{eq:d_inst_bh} are linear combinations of
the doubled instanton equations defined by $\mathcal{J}^B_{J_{\pm}}$ 
and $\mathcal{J}^B_{\omega_{\pm}}$.
As we have clarified, under the chirality condition, we have the
following equations from the doubled instanton equations;
\begin{gather}
\begin{aligned}
{\dop} X^{\mu} \pm (J_+)^{\mu} {}_{\nu} * {\dop} X^{\nu} &= 0,
\\
{\dop} X^{\mu} \pm (J_-)^{\mu} {}_{\nu} * {\dop} X^{\nu} &= 0,
\notag
\end{aligned}
\\
\begin{align}
(J_+)^{\mu} {}_{\nu} {\dop} X^{\nu} &= \mp i {\dop} X^{\mu},
\notag \\
(J_-)^{\mu} {}_{\nu} {\dop} X^{\nu} &= \mp i {\dop} X^{\mu}.
\end{align}
\end{gather}
This means that the solutions are restricted to the (anti)holomorphic
vectors defined by $J_{\pm}$ and they are also instantons with respect
to $J_{\pm}$. This is possible since the bi-hermitian structure
satisfies $[J_+,J_-] = 0$ and the common eigenvectors of $J_{\pm}$ are allowed.
This also implies $J_{\pm} *$ have common eigenvectors (instantons).
The action bound in this case is
\begin{align}
S_{\text{E}} =& \  \frac{1}{2} \int g_{\mu \nu} {\dop} X^{\mu} \wedge * {\dop} X^{\nu} 
+ \frac{i}{2} \int B_{\mu \nu} {\dop} X^{\mu} \wedge {\dop} X^{\nu}
\notag \\
=& \ 
\pm \frac{1}{2} \int (\omega_+)_{\mu \nu} {\dop} X^{\mu} \wedge {\dop} X^{\nu} 
+ \frac{i}{2} \int B_{\mu \nu} {\dop} X^{\mu} \wedge {\dop} X^{\nu}
\notag \\
=& \ 
\pm \frac{1}{2} \int (\omega_-)_{\mu \nu} {\dop} X^{\mu} \wedge {\dop} X^{\nu} 
+ \frac{i}{2} \int B_{\mu \nu} {\dop} X^{\mu} \wedge {\dop} X^{\nu}.
\end{align}
This is nothing but the bound for the ordinary worldsheet instantons
defined by $J_{\pm}$.

\section{Conclusion and discussions} \label{sec:conclusion}
In this paper, we studied doubled structures that encode K\"{a}hler,
hyperk\"{a}hler, bi-hermitian and bi-hypercomplex geometries.

The spacetime metric $g_{\mu \nu}$ and the NSNS $B$-field are organized into the generalized
metric $\mathcal{H}_{MN}$ and the natural $O(D,D)$ structures of DFT are implemented by
the Born structure on the doubled space $\mathcal{M}$.
Due to the natural isomorphism emerged from the Born structure, the tangent
bundle of the doubled space $\mathcal{M}$ and the generalized tangent
bundle $\mathbb{T}M$ over the physical spacetime $M$ is identified.
On the other hand, the K\"{a}hler structure on the physical spacetime $M$ is embedded
into the generalized K\"{a}hler structure $(\mathcal{J}_{J}, \mathcal{J}_{\omega})$ on $T \mathcal{M}$.
We analyzed compatibility 
of the doubled and the Born structures in the doubled space.
We found that the algebraic structures require the extra doubled structures 
$\mathcal{P}$ and $\mathcal{Q}$ in the Born geometry.
Altogether we showed that they form the algebra of the bi-quaternions.
The Born and the generalized complex structures appear as subalgebras of
split-quaternions and the bi-complex numbers, respectively.
Utilizing this fact, we extended the discussion to the bi-hermitian case.
We found that the desired algebra that encodes the bi-hermitian
structure on spacetime is the algebra of bi-quaternions over $\mathbb{C}$.
This is a 16-dimensional algebra that contains appropriate subalgebras.
By using the basis of the algebra, we write down all the real and
imaginary units in their explicit forms.
For the hyperk\"{a}hler structure on spacetime, 
it is represented by the generalized hyperk\"{a}hler structure on $T\mathcal{M}$.
This satisfies the algebra of split-bi-quaternions.
This together with the Born structure leads us to the algebra of
split-tetra-quaternions.
We exhibited the explicit representations of the doubled structures that
form this algebra.
We further extended the results to 
the bi-hypercomplex case.
We found that the structure is realized as the algebra of the
split-tetra-quaternions over $\mathbb{H}$ in the doubled space.
These results provide us deep connections among
the algebras of the hypercomplex numbers, the complex structures of 
spacetime, the doubled structures and T-duality.
We also showed that some of the algebras of the hypercomplex numbers
also expressed by Clifford algebras.

In the latter part of this paper, we studied the doubled worldsheet
instantons in the Born sigma model.
The Born sigma model is a sigma model whose target space is the Born
geometry.
The model keeps manifest T-duality and is governed by the generalized
metric $\mathcal{H}_{MN}$ and the topological term.
The ordinary string sigma model is reproduced by the DFT constraints
and the chirality condition defined by $\mathcal{J}$.
We derived the Bogomol'nyi equations defined by the doubled complex structures on $T \mathcal{M}$.
We clarify that appropriate doubled complex structures reproduced the
ordinary worldsheet instanton equations.
We then discussed the T-duality transformation of the worldsheet
instantons. We in particular focused on the T-duality between 
K\"{a}hler and bi-hermitian geometries.
The one-to-two correspondence of the worldsheet instantons discussed in 
\cite{Kimura:2022dma} is naturally interpreted in the Born sigma model.
We showed that the instantons in the bi-hermitian geometries are
represented by a linear combination of individual instantons defined by
the structures $(J_{\pm}, \omega_{\pm})$.
The analysis can be extended to the hyperk\"{a}hler and bi-hypercomplex cases.
The Bogomol'nyi equations in the Born sigma models are interpreted as 
the T-duality covariant realization of the worldsheet instanton equations.

As we discussed, the doubled space plays an important role 
in revealing the T-duality among geometric structures.
It has been discussed that solutions to supergravities that are related
by T-duality transformations are given by a solution to DFT.
This means that the spacetime geometries of various supergravity solutions
are described by doubled geometry in a T-duality unified manner.
For example, the H-monopole (smeared NS5-brane) and the KK-monopole
(KK5-brane) in type II string theories are 
unified into an $O(D,D)$ covariant solution to DFT \cite{Berman:2014jsa}.
The worldsheet instanton effects in the H-monopole geometry, the
KK-monopole geometry and their relations to T-duality are studied in
various perspectives \cite{Gregory:1997te, Tong:2002rq, Harvey:2005ab,
Okuyama:2005gx}.
Among other things, the instantons break the isometry of the H-monopole
geometry and recover that of the NS5-brane which is a genuine solution
in string theory.
Things get more interesting when we consider this phenomenon in the
T-dual side. It has been shown that instantons in the KK-monopole
geometry breaks the isometry not along the KK direction, but of the
winding space \cite{Harvey:2005ab,Okuyama:2005gx}.
The modified geometry is characterized not only by the physical
coordinate $x^{\mu}$ but also by the winding coordinate
$\tilde{x}_{\mu}$.
These fact mean that instantons reveal the more stringy nature of
spacetime.
It would be interesting to study this winding geometry in the context of
the Born sigma model.
The notion of T-duality covariant instantons 
helps us to understand geometries that are not fully
captured in supergravities. They are known as non-geometries.
An example of this kind of non-geometry is the T-fold \cite{Hull:2004in,
Hull:2006va} whose explicit realization 
includes exotic branes in string theories \cite{deBoer:2010ud}.
The exotic $5^2_2$-brane in type II string theories 
is a typical example studied intensively.
Indeed, the exotic $5^2_2$-brane obtained by the
T-duality transformation of the hyperk\"{a}hler (Taub-NUT) geometry 
is realized by a solution to DFT in a specific frame \cite{Bakhmatov:2016kfn}.
It has been shown that the $5^2_2$-brane geometry is expected to admit
the bi-hypercomplex structures \cite{Kimura:2022dma} and the worldsheet instantons in the
$5^2_2$-brane geometry is studied \cite{Kimura:2013fda, Kimura:2013zva, Kimura:2018hph}.
It would be interesting to study the instantons and bi-hypercomplex
structures in the doubled setup.
We will come back to these issues in future researches.

\subsection*{Acknowledgments}
We would like to thank Michael J. Duff and Ulf Lindstr\"{o}m for correspondence.
The work of T.K. and S.S. is supported in part by Grant-in-Aid for Scientific Research (C),
JSPS KAKENHI Grant Number JP20K03952. The work of K.S. is supported by Grant-in-Aid
for JSPS Research Fellow, JSPS KAKENHI Grant Number JP20J13957.


\begin{appendix}

\section{Mathematics on hypercomplex numbers} \label{sec:appendix1}
In this appendix, we provide a brief introduction of hypercomplex
numbers.
The materials here are the minimum definitions and properties required
to understand the main text.
Readers who need mathematically rigorous definitions would consult literature.

\subsection{Basic elements}
\paragraph{Binarions.}
Binarions are the two-dimensional 
(non)associative unital
algebras over
the field $\mathbb{R}$.
Binarions are classified as the followings depending on the bases of the algebras.
\begin{quote}
\begin{enumerate}
\item Complex numbers are generated by the basis $(1,i)$; \ $i^2 = -1$.
\item Split-complex numbers are generated by the basis $(1,j)$; \ $j^2 = 1$.
\item Dual numbers are generated by the basis $(1,\varepsilon)$; \ $\varepsilon^2 = 0$.
\end{enumerate}
\end{quote}
Note that the complex numbers define the field $\mathbb{C}$ but the
split-complex and dual numbers do not. This is because they have
non-trivial zero divisors.
Since we never treat the dual numbers in this paper, we do not care
about the ``dual''-hypercomplex numbers. 

\paragraph{Quaternions.}
We next introduce four dimensional 
(non)associative unital 
algebras over the field $\mathbb{R}$.
There are two options.
\begin{quote}
\begin{enumerate}
\item Quaternions are defined by a normed (associative) division algebra over
the field $\mathbb{R}$. This is defined by the basis $(1,i,j,k)$ given
      in the product Table \ref{tb:quaternion} (left).
\item Split-quaternions are defined by the basis $(1,i,j,k)$ given in
      the product Table \ref{tb:quaternion} (right).
\end{enumerate}
\end{quote}
The quaternions define a field $\mathbb{H}$ but the split-quaternions do not.

\begin{table}[t]
\centering
\begin{tabular}{c||c|c|c}
 & $i$ & $j$ & $k$ \\
\hline
\hline
$i$ & $-1$& $k$ & $-j$ \\
\hline
$j$ & $-k$ & $-1$ & $i$ \\
\hline
$k$ & $j$ & $-i$ & $-1$
\end{tabular}
\qquad
\begin{tabular}{c||c|c|c}
 & $i$ & $j$ & $k$ \\
\hline
\hline
$i$ & $-1$& $k$ & $-j$ \\
\hline
$j$ & $-k$ & $1$ & $-i$ \\
\hline
$k$ & $j$ & $i$ & $1$
\end{tabular}
\caption{
The product tables of the bases of quaternions (left) and split-quaternions (right).
For quaternions $i^2 = j^2 = k^2 = ijk = -1$ and they anti-commute.
For split-quaternions $i^2 = -1, j^2 = k^2 = ijk = 1$ and they anti-commute.
The split-quaternions are obtained by replacing $j \to \mathbf{i}j, k \to
 \mathbf{i} k$ in the quaternions. Here $\mathbf{i}$ is an auxiliary
 imaginary unit $\mathbf{i}^2 = -1$.
}
\label{tb:quaternion}
\end{table}

\subsection{Unital division algebras over fields $\mathbb{C}$ and $\mathbb{H}$}
Some hypercomplex numbers over the field $\mathbb{R}$ 
are defined as unital division algebras over the fields $\mathbb{C}$ and 
$\mathbb{H}$.
The relevant examples are the 
followings;
\begin{quote}
\begin{enumerate}
\item $\mathbb{C}$ over $\mathbb{C}$ -- bi-complex numbers,
\item $\mathrm{Sp}\mathbb{C}$ over $\mathbb{C}$ -- split-bi-complex numbers,
\item $\mathbb{C}$ over $\mathbb{H}$ -- bi-quaternions,
\item $\mathrm{Sp}\mathbb{C}$ over $\mathbb{H}$ -- split-bi-quaternions,
\item $\mathbb{H}$ over $\mathbb{H}$ -- tetra-quaternions,
\item $\mathrm{Sp}\mathbb{H}$ over $\mathbb{H}$ -- split-tetra-quaternions.
\end{enumerate}
\end{quote}
Here $\mathrm{Sp}\mathbb{C}$ and 
$\mathrm{Sp}\mathbb{H}$ stand for split-complex numbers and split-quaternions, respectively.
They are schematically represented by tensor products of the fields.
For example, the bi-complex numbers $\mathbb{C}_2$ are identified with $\mathbb{C}
\otimes \mathbb{C}$.
The algebra $\mathbb{H}$ over $\mathbb{C}$ and $\mathrm{Sp}\mathbb{H}$
over $\mathbb{C}$ are isomorphic to bi-quaternions and
split-bi-quaternions, respectively.

\paragraph{Bi-complex numbers.}
The bi-complex numbers are defined by complex numbers over the field $\mathbb{C}$.
For $x,y \in \mathbb{C}$, a bi-complex number $X$ is represented by
\begin{align}
X = x \mathbf{1} + y \mathbf{i},
\end{align}
where $\mathbf{1}^2 = \mathbf{1}, \mathbf{i}^2 = -
\mathbf{1}$ are bases of the complex numbers.
Since the coefficients $x,y$ are expanded by the basis of the complex
numbers $(1,i)$ with real coefficients, the basis of the bi-complex
numbers is given by
\begin{align}
e_0 = 1 \mathbf{1}, 
\quad
e_1 = 1 \mathbf{i}, 
\quad
e_2 = i \mathbf{1}, 
\quad
e_3 = i \mathbf{i}.
\end{align}
All the quantities $1,i,\mathbf{1}, \mathbf{i}$ commute with each other.
Then, we have the algebra
\begin{gather}
e_0^2 = e_0, 
\quad 
e_1^2 = - e_0, 
\quad 
e_2^2 = - e_0, 
\quad 
e_3^2 = e_0,
\notag \\
e_1 e_2 = e_3,
\quad
e_2 e_3 = - e_1,
\quad 
e_3 e_1 = - e_2,
\quad 
e_1 e_2 e_3 = e_0.
\end{gather}
We have two real units $e_0, e_3$ and two imaginary units $e_1, e_2$.
The algebra defines the product table of the basis Table
\ref{tb:bi-complex}.
The bi-complex numbers are also known as tessarine.

Similarly, we can consider bi-complex numbers over $\mathbb{C}$.
This is known as the tri-complex numbers by Segre \cite{Segre}.
The basis of the tri-complex numbers is
$(1 \mathbf{1} \hat{1}, i \mathbf{i} \hat{1}, 1 \mathbf{i} \hat{i}, i
\mathbf{1} \hat{i}, 
1 \mathbf{i} \hat{1}, i \mathbf{1} \hat{1}, 1 \mathbf{1} \hat{i}, i \mathbf{i}
\hat{i})$ 
where $(\hat{1}, \hat{i})$ is
the additional basis
of the complex numbers.

\begin{table}[t]
\centering
\begin{tabular}{c||c|c|c}
 & $i$ & $j$ & $k$ \\
\hline
\hline
$i$ & $-1$& $k$ & $-j$ \\
\hline
$j$ & $k$ & $-1$ & $-i$ \\
\hline
$k$ & $-j$ & $-i$ & $1$
\end{tabular}
\caption{The product table for the bi-complex numbers.
The bases of the bi-complex numbers $i^2 = j^2 - 1, k^2 = 1, ijk=1$ all commute.
}
\label{tb:bi-complex}
\end{table}

\paragraph{Split-bi-complex numbers.}
The split-bi-complex numbers are split-complex numbers over the field $\mathbb{C}$.
The basis of the split-bi-complex numbers is
\begin{align}
e_0 = 1 \mathbf{1}, \quad e_1 = i \mathbf{1}, \quad e_2 = 1 \mathbf{j}, \quad e_3 = i \mathbf{j},
\end{align}
where $(1,i)$ and $(\mathbf{1}, \mathbf{j})$ are the bases of the complex
and the split-complex numbers.
They satisfy
\begin{gather}
e_0^2 = e_0, 
\quad 
e_1^2 = - e_0, 
\quad 
e_2^2 = e_0, 
\quad 
e_3^2 = - e_0,
\notag \\
e_1 e_2 = e_3,
\quad 
e_2 e_3 =  e_1,
\quad 
e_3 e_1 = - e_2,
\quad 
e_1 e_2 e_3 = - e_0.
\end{gather}
We find that if we redefine
\begin{align}
e_0' = e_0, \quad 
e_1' = e_1, \quad
e_2' = - e_3, \quad 
e_3' = e_2,
\end{align}
the algebra becomes that of the bi-complex numbers.
Therefore the bi-complex and split-bi-complex numbers are isomorphic
with each other.

\paragraph{Bi-quaternions.}
The bi-quaternions over the field $\mathbb{R}$ are defined as quaternions over the field $\mathbb{C}$,
or equivalently, complex numbers over the field $\mathbb{H}$.
The basis is given by
\begin{align}
e_0 \mathbf{1}, \ e_1 \mathbf{1}, \ e_2 \mathbf{1}, \ e_3 \mathbf{1}, \ 
e_0 \mathbf{i}, \ e_1 \mathbf{i}, \ e_2 \mathbf{i}, \ e_3 \mathbf{i},
\end{align}
where $e_{\mu} \, (\mu=0,1,2,3)$ and $(\mathbf{1}, \mathbf{i})$ are
bases of the quaternions and the complex numbers.
The bi-quaternion algebra is associative, non-commutative and normed.
By using the quaternion algebra $(e_0, e_1, e_2, e_3)$, we have the relations
\begin{align}
(e_i \mathbf{1}) (e_j \mathbf{1}) 
&= - \delta_{ij} (e_0 \mathbf{1}) + \epsilon_{ijk} (e_k \mathbf{1}),
\notag \\
(e_i \mathbf{i})(e_j \mathbf{i}) 
&= + \delta_{ij} (e_0 \mathbf{1}) - \epsilon_{ijk} (e_k \mathbf{1}),
\notag \\
(e_i \mathbf{1}) (e_j \mathbf{i}) 
&= - \delta_{ij} (e_0 \mathbf{i}) + \epsilon_{ijk} (e_k \mathbf{i}),
\notag \\
(e_i \mathbf{i}) (e_j \mathbf{1}) 
&= - \delta_{ij} (e_0 \mathbf{i}) + \epsilon_{ijk} (e_k \mathbf{i}),
\end{align}
When we define $a_i = e_i \mathbf{1}, b_i = e_i \mathbf{i}, c =
e_0 \mathbf{i}, 1 = e_0 \mathbf{1}$, 
they satisfy
\begin{gather}
\begin{aligned}
a_i a_j &= - \delta_{ij} 1 + \epsilon_{ijk} a_k,
\notag \\
b_i b_j &= \delta_{ij} 1 - \epsilon_{ijk} a_k,
\notag \\
a_i b_j &= - \delta_{ij} c + \epsilon_{ijk} b_k,
\notag \\
b_i a_j &= - \delta_{ij} c + \epsilon_{ijk} b_k,
\notag 
\end{aligned}
\\
a_i^2 = c^2 = -1, \ b_i^2 = 1^2 = + 1, \ (i=1,2,3).
\end{gather}
This is the algebra that the bi-quaternions satisfy.
The algebra involves the following subalgebras;
\begin{quote}
\begin{enumerate}
\item $(e_0 \mathbf{i}, e_1 \mathbf{1}, e_1 \mathbf{i})$, 
      $(e_0 \mathbf{i}, e_2 \mathbf{1}, e_2 \mathbf{i})$, 
      $(e_0 \mathbf{i}, e_3 \mathbf{1}, e_3 \mathbf{i})$ :
      bi-complex numbers,
\item $(e_1 \mathbf{1}, e_2 \mathbf{i}, e_3 \mathbf{i})$, 
      $(e_1 \mathbf{i}, e_2 \mathbf{1}, e_3 \mathbf{i})$, 
      $(e_1 \mathbf{i}, e_2 \mathbf{i}, e_3 \mathbf{1})$ : 
      split-quaternions,
\item $(e_1 \mathbf{1}, e_2 \mathbf{1}, e_3 \mathbf{1})$ : 
     quaternions.
\end{enumerate}
\end{quote}

\paragraph{Bi-quaternions over the field $\mathbb{C}$.}
Bi-quaternions over the field $\mathbb{C}$ are defined by the basis
\begin{align}
&
e_0 \mathbf{1} \hat{1},
\quad
e_i \mathbf{1} \hat{1},
\quad
e_0 \mathbf{i} \hat{1},
\quad
e_i \mathbf{i} \hat{1},
\quad 
e_0 \mathbf{1} \hat{i},
\quad
e_i \mathbf{1} \hat{i},
\quad
e_0 \mathbf{i} \hat{i},
\quad
e_i \mathbf{i} \hat{i},
\end{align}
where $(\hat{1}, \hat{i})$ is the additional basis of complex numbers.
This defines a 16-dimensional algebra involving 8 real and 8 imaginary units.
The algebra contains the bi-complex numbers as subalgebras;
\begin{align}
(e_0 \mathbf{i} \hat{1}, e_i \mathbf{1} \hat{1}, e_i \mathbf{i}
 \hat{1}),
\qquad
(e_0 \mathbf{1} \hat{i}, e_i \mathbf{i} \hat{i}, e_i \mathbf{i} \hat{1}).
\end{align}

\paragraph{Split-bi-quaternions.}
The split-bi-quaternions are split-complex numbers over the field $\mathbb{H}$.
The basis is
\begin{align}
e_0 \mathbf{1}, \ \ e_1 \mathbf{1}, \ \ e_2 \mathbf{1}, \ \ e_3 \mathbf{1}, 
\ \ e_0 \TS \mathbf{j}, 
\ \ e_1 \TS \mathbf{j}, 
\ \ e_2 \TS \mathbf{j}, 
\ \ e_3 \TS \mathbf{j},
\end{align}
where $e_{\mu} \, (\mu = 0,1,2,3)$ and $(1,\mathbf{j})$ are bases of the
quaternions and the split-complex numbers.
The basis of the split-bi-quaternions satisfies
\begin{gather}
(e_0 \mathbf{1})^2 =  e_0 \mathbf{1}, \qquad
(e_{i} \mathbf{1})^2 = - e_0 \mathbf{1}, \qquad 
(e_0 \TS \mathbf{j})^2 =  e_0 \mathbf{1}, \qquad
(e_i \TS \mathbf{j})^2 = - e_0 \mathbf{1},
\notag \\
(e_1 \mathbf{1}) (e_2 \mathbf{1}) (e_3 \mathbf{1})
(e_0 \TS \mathbf{j}) (e_1 \TS \mathbf{j})
(e_2 \TS \mathbf{j}) (e_3 \TS \mathbf{j}) = e_0 \mathbf{1}.
\end{gather}
Since we have 
\begin{align}
(e_i \mathbf{1}) (e_j \mathbf{1}) 
&= - \delta_{ij} (e_0 \mathbf{1}) + \epsilon_{ijk} (e_k \mathbf{1}),
\notag \\
(e_i \, \mathbf{j})(e_j \TS \mathbf{j}) 
&= - \delta_{ij} (e_0 \mathbf{1}) + \epsilon_{ijk} (e_k \mathbf{1}),
\notag \\
(e_i \mathbf{1})  (e_j \TS \mathbf{j}) 
&= - \delta_{ij} (e_0 \TS \mathbf{j}) + \epsilon_{ijk} (e_k \TS \mathbf{j}),
\notag \\
(e_i \, \mathbf{j}) (e_j \mathbf{1}) 
&= - \delta_{ij} (e_0 \TS \mathbf{j}) + \epsilon_{ijk} (e_k \TS \mathbf{j}),
\end{align}
if we define
\begin{align}
\mathcal{J}_{i,+} = e_i \mathbf{1}, \qquad 
\mathcal{J}_{i,-} = e_i \TS \mathbf{j}, \qquad
\mathcal{G} = - e_0 \TS \mathbf{j}, \qquad 
\mathbf{1}_{2d} = e_0 \mathbf{1},
\end{align}
they satisfy $\mathcal{J}_{i,+}^2 = - \mathbf{1}_{2d}$,
$\mathcal{J}_{i,-}^2 = - \mathbf{1}_{2d}$, $\mathcal{G}^2 = \mathbf{1}_{2d}$ and
\begin{alignat}{2}
\mathcal{J}_{i,+} \mathcal{J}_{j,+}
&= - \delta_{ij} \mathbf{1}_{2d} + \epsilon_{ijk} \mathcal{J}_{k,+},
&\qquad 
\mathcal{J}_{i,-} \mathcal{J}_{j,-}
&= - \delta_{ij} \mathbf{1}_{2d} + \epsilon_{ijk} \mathcal{J}_{k,+},
\notag \\
\mathcal{J}_{i,+} \mathcal{J}_{j,-}
&= 
\delta_{ij} \mathcal{G} + \epsilon_{ijk} \mathcal{J}_{k,-},
&\qquad
\mathcal{J}_{i,-} \mathcal{J}_{j,+}
&= 
 \delta_{ij} \mathcal{G} + \epsilon_{ijk} \mathcal{J}_{k,-}.
\end{alignat}
This is the algebra of the generalized hyperk\"{a}hler structure \eqref{eq:generalized_hyperkahler_algebra}.

The subalgebras of the split-bi-quaternions are the followings;
\begin{quote}
\begin{enumerate}
\item $(e_0 \TS \mathbf{j}, e_{1} \mathbf{1}, e_1 \TS \mathbf{j})$, 
      $(e_0 \TS \mathbf{j}, e_{2} \mathbf{1}, e_2 \TS \mathbf{j})$, 
      $(e_0 \TS \mathbf{j}, e_{3} \mathbf{1}, e_3 \TS \mathbf{j})$ : 
      bi-complex numbers,
\item $(e_1 \mathbf{1}, e_{2} \TS \mathbf{j}, e_3 \TS \mathbf{j})$, 
      $(e_1 \TS \mathbf{j}, e_{2} \mathbf{1}, e_3 \TS \mathbf{j})$, 
      $(e_1 \TS \mathbf{j}, e_{2} \TS \mathbf{j}, e_3 \mathbf{1})$, 
      $(e_1 \mathbf{1}, e_2 \mathbf{1}, e_3 \mathbf{1})$ : 
      quaternions.
\end{enumerate}
\end{quote}

\paragraph{Split-bi-quaternions over $\mathbb{H}$.}
Split-bi-quaternions over $\mathbb{H}$ are 
defined by the basis;
\begin{alignat}{4}
&
e_0 \mathbf{1} \hat{e}_0,
&\qquad
&e_0 \mathbf{1} \hat{e}_i,
&\qquad
&e_i \mathbf{1} \hat{e}_0,
&\qquad
&e_i \mathbf{1} \hat{e}_j,
\notag \\
&
e_0 \TS \mathbf{j} \hat{e}_0,
&\qquad
&e_0 \TS \mathbf{j} \hat{e}_i,
&\qquad
&e_i \TS \mathbf{j} \hat{e}_0,
&\qquad
&e_i \TS \mathbf{j} \hat{e}_j.
\label{eq:split-bi-quaternions-H}
\end{alignat}
Here $e_{\mu}$ and $\hat{e}_{\mu}$ are the bases of two commuting
quaternions and $(\mathbf{1}, \mathbf{j})$ is the basis of split-complex numbers.
The basis \eqref{eq:split-bi-quaternions-H} defines a 32-dimensional
algebra involving 20 real and 12 imaginary units.
The subalgebra involves six bi-complex numbers that share one real unit.
They are easily constructed as 
\begin{align}
(e_0 \TS \mathbf{j} \hat{e}_0, e_a \mathbf{1} \hat{e}_0, e_a \TS \mathbf{j}
 \hat{e}_0),
\qquad
(e_0 \TS \mathbf{j} \hat{e}_0, e_0 \mathbf{1} \hat{e}_a, e_0 \TS \mathbf{j}
 \hat{e}_a),
\qquad 
(a=1,2,3).
\end{align}
Here the common real unit is $e_0 \TS \mathbf{j} \hat{e}_0$.

\paragraph{Tetra-quaternions.}
The tetra-quaternions are quaternions over the field $\mathbb{H}$ \cite{Girard:2007}.
The basis is given by
\begin{align}
e_{\mu} \mathbf{e}_{\nu}, \ (\mu,\nu = 0,1,2,3).
\end{align}
There are 6 imaginary units $e_0 \mathbf{e}_i, e_i \mathbf{e}_0 \,
(i=1,2,3)$ and 10 real units $e_0 \mathbf{e}_0, e_i \mathbf{e}_j$.
In the following, we denote $e_0 = 1, \mathbf{e}_0 = \mathbf{1}$, $1 \mathbf{e}_{\mu} = \mathbf{e}_{\mu}$.
The algebra contains the following subalgebras;
\begin{quote}
\begin{enumerate}
\item $(e_i \mathbf{1}, \mathbf{e}_1, e_i \mathbf{e}_1)$,
      $(e_i \mathbf{1}, \mathbf{e}_2, e_i \mathbf{e}_2)$,
      $(e_i \mathbf{1}, \mathbf{e}_3, e_i \mathbf{e}_3)$, $(i=1,2,3)$ : 
      bi-complex numbers,
\item $(\mathbf{e}_1, \mathbf{e}_2, \mathbf{e}_3)$, 
      $(e_1 \mathbf{1}, e_2 \mathbf{1}, e_3 \mathbf{1})$ : 
      quaternions,
\item $(\mathbf{e}_1, e_i \mathbf{e}_2, e_i \mathbf{e}_3)$, 
      $(e_i \mathbf{e}_1, \mathbf{e}_2, e_i \mathbf{e}_3)$, 
      $(e_i \mathbf{e}_1, e_i \mathbf{e}_2, \mathbf{e}_3)$,
      $(e_1 \mathbf{1}, e_2 \mathbf{e}_i, e_3 \mathbf{e}_i)$, \\
      $(e_1 \mathbf{e}_i, e_2 \mathbf{1}, e_3 \mathbf{e}_i)$, 
      $(e_1 \mathbf{e}_i, e_2 \mathbf{e}_i, e_3 \mathbf{1})$, $(i=1,2,3)$ : 
      split-quaternions,
\item $(\mathbf{e}_{\mu}, e_i \mathbf{e}_{\mu}), \, (i=1,2,3)$ : 
      bi-quaternions.
\end{enumerate}
\end{quote}
Note that the tetra-quaternion algebra contains two commuting quaternions.
This reflects the property of the bi-hypercomplex structures.

\paragraph{Split-tetra-quaternions.}
The split-tetra-quaternions are split-quaternions over the field $\mathbb{H}$.
The basis is as follows;
\begin{alignat}{4}
&
e_0 \mathbf{e}_0, 
&\qquad 
&e_0 \mathbf{e}_i, 
&\qquad 
&e_1 \mathbf{e}_0,
&\qquad
&e_1 \mathbf{e}_i, 
\notag \\
&
i e_2 \mathbf{e}_0,
&\qquad
&i e_2 \mathbf{e}_i, 
&\qquad
&i e_3 \mathbf{e}_0,
&\qquad 
&i e_3 \mathbf{e}_i.
\label{eq:basis_stq}
\end{alignat}
They satisfy
\begin{gather}
(e_0 \mathbf{e}_0)^2 = e_0 \mathbf{e}_0,
\quad
(e_0 \mathbf{e}_i)^2 = - e_0 \mathbf{e}_0,
\quad 
(e_1 \mathbf{e}_0)^2 = - e_0 \mathbf{e}_0,
\quad
(e_1 \mathbf{e}_i)^2 = e_0 \mathbf{e}_0, 
\notag \\
(i e_2 \mathbf{e}_0)^2 = e_0 \mathbf{e}_0,
\quad
(i e_2 \mathbf{e}_i)^2 = - e_0 \mathbf{e}_0, 
\quad
(i e_3 \mathbf{e}_0)^2 = e_0 \mathbf{e}_0,
\quad 
(i e_3 \mathbf{e}_i)^2 = - e_0 \mathbf{e}_0,
\end{gather}
and involve 6 real and 10 imaginary units.
The subalgebras are the followings;
\begin{quote}
\begin{enumerate}
\item $(e_1 \mathbf{1}, \mathbf{e}_1, e_1 \mathbf{e}_1)$, 
      $(i e_{2} \mathbf{1}, \mathbf{e}_1, i e_{2} \mathbf{e}_1) $,
      $(i e_{3} \mathbf{1}, \mathbf{e}_1, i e_{3} \mathbf{e}_1) $,
      $(e_1 \mathbf{1}, \mathbf{e}_2, e_1 \mathbf{e}_2)$,
      $(i e_{2} \mathbf{1}, \mathbf{e}_2, i e_{2} \mathbf{e}_2)$,
      $(i e_{3} \mathbf{1}, \mathbf{e}_2, i e_{3} \mathbf{e}_2)$,
      $(e_1 \mathbf{1}, \mathbf{e}_3, e_1 \mathbf{e}_3)$,
      $(i e_{2} \mathbf{1}, \mathbf{e}_3, i e_{2} \mathbf{e}_3)$,
      $(i e_{3} \mathbf{1}, \mathbf{e}_3, i e_{3} \mathbf{e}_3)$ :
      bi-complex numbers,
\item $(\mathbf{e}_1, \mathbf{e}_2, \mathbf{e}_3)$,
      $(\mathbf{e}_1, i e_{2} \mathbf{e}_2, i e_{2} \mathbf{e}_3)$,
      $(\mathbf{e}_1, i e_{3} \mathbf{e}_2, i e_{3} \mathbf{e}_3)$,
      $(i e_{2} \mathbf{e}_1, \mathbf{e}_2, i e_{2} \mathbf{e}_3)$,
      $(i e_{3} \mathbf{e}_1, \mathbf{e}_2, i e_{3} \mathbf{e}_3)$,
      $(i e_{2} \mathbf{e}_1, i e_{2} \mathbf{e}_2, \mathbf{e}_3)$,
      $(i e_{3} \mathbf{e}_1, i e_{3} \mathbf{e}_2, \mathbf{e}_3)$,
      $(e_1 \mathbf{1}, i e_2 \mathbf{e}_i, i e_3 \mathbf{e}_i)$,
      $(i=1,2,3)$ : quaternions,
\item $(\mathbf{e}_1, e_1 \mathbf{e}_2, e_1 \mathbf{e}_3)$,
      $(e_1 \mathbf{e}_1, \mathbf{e}_2, e_1 \mathbf{e}_3)$,
      $(e_1 \mathbf{e}_1, e_1 \mathbf{e}_2, \mathbf{e}_3)$,
      $(e_1 \mathbf{1}, i e_2 \mathbf{1}, i e_3 \mathbf{1})$,
      $(e_1 \mathbf{1}, e_2 \mathbf{e}_i, e_3 \mathbf{e}_i)$,
      $(e_1 \mathbf{e}_i, i e_2 \mathbf{1}, i e_3 \mathbf{e}_i)$,
      $(e_1 \mathbf{e}_i, i e_2 \mathbf{e}_i, i e_3 \mathbf{1})$, $(i=1,2,3)$
      : split-quaternions,
\item $(\mathbf{e}_{\mu}, e_1 \mathbf{e}_{\mu})$ : bi-quaternions,
\item $(\mathbf{e}_{\mu}, i e_{2} \mathbf{e}_{\mu})$ : split-bi-quaternions,
\item $(\mathbf{e}_{\mu}, i e_{3} \mathbf{e}_{\mu})$ : split-bi-quaternions.
\end{enumerate}
\end{quote}
Here we have denoted $e_0 \mathbf{e}_{\mu}$ as $\mathbf{e}_{\mu}$.
The split-tetra-quaternions contain split-bi-quaternions as subalgebras.
This contains algebras of the generalized hyperk\"{a}hler structures
and the Born structures.

\paragraph{Split-tetra-quaternions over the field $\mathbb{H}$.}
The basis of split-tetra-quaternions over the field $\mathbb{H}$ is 
\begin{alignat}{8}
&
e_0 \mathbf{e}_0 \hat{\mathbf{e}}_0, 
&\quad 
&e_0 \mathbf{e}_i \hat{\mathbf{e}}_0,
&\quad
&e_1 \mathbf{e}_0 \hat{\mathbf{e}}_0,
&\quad
&e_1 \mathbf{e}_i \hat{\mathbf{e}}_0,
&\quad
&i e_2 \mathbf{e}_0 \hat{\mathbf{e}}_0,
&\quad
&i e_2 \mathbf{e}_i \hat{\mathbf{e}}_0,
&\quad
&i e_3 \mathbf{e}_0 \hat{\mathbf{e}}_0,
&\quad
&i e_3 \mathbf{e}_i \hat{\mathbf{e}}_0,
\notag \\
&
e_0 \mathbf{e}_0 \hat{\mathbf{e}}_i,
&\quad 
&e_0 \mathbf{e}_i \hat{\mathbf{e}}_j,
&\quad
&e_1 \mathbf{e}_0 \hat{\mathbf{e}}_i,
&\quad
&e_1 \mathbf{e}_i \hat{\mathbf{e}}_j,
&\quad
&i e_2 \mathbf{e}_0 \hat{\mathbf{e}}_i,
&\quad 
&i e_2 \mathbf{e}_i \hat{\mathbf{e}}_j,
&\quad
&i e_3 \mathbf{e}_0 \hat{\mathbf{e}}_i,
&\quad
&i e_3 \mathbf{e}_i \hat{\mathbf{e}}_j.
\end{alignat}
Here $e_{\mu}$, $\mathbf{e}_{\mu}$ and $\hat{e}_{\mu}$ are quaternions
that commute with each other. The basis defines a
64-dimensional algebra involving
36 real and 28 imaginary units.
The algebra contains split-tetra-quaternions as a subalgebra and 
6 bi-complex numbers that share one real unit $i e_2 \mathbf{e}_0 \hat{\mathbf{e}}_0$;
\begin{align}
(i e_2 \mathbf{e}_0 \hat{\mathbf{e}}_0, e_0 \mathbf{e}_i
 \hat{\mathbf{e}}_0, i e_2 \mathbf{e}_i \hat{\mathbf{e}}_0),
\qquad 
(i e_2 \mathbf{e}_0 \hat{\mathbf{e}}_0, e_0 \mathbf{e}_0
 \hat{\mathbf{e}}_i, i e_2 \mathbf{e}_0 \hat{\mathbf{e}}_i),
\qquad 
(i=1,2,3).
\end{align}

\section{Clifford algebra and hypercomplex numbers} \label{sec:appendix2}
Some hypercomplex numbers are related to Clifford algebras.
In this appendix, we present the explicit relations among them.
We first define quantities
\begin{alignat}{2}
e_1^2 = e_2^2 = \cdots = e_p^2 &= 1, &\qquad & 
\notag \\
e_{p+1}^2 = \cdots = e_n^2 &= - 1, &\qquad &
\notag \\
e_i e_j + e_j e_i &= 0, &\qquad &(i \neq j).
\end{alignat}
A Clifford algebra $Cl_{p,q} (\mathbb{R})$ is defined by the basis
$1, e_i, e_i \wedge e_j, e_i \wedge e_j \wedge e_k, \cdots$.
Here $1$ is the unit of the multiplication of the field $\mathbb{R}$.
In the following, we use $e_i \wedge e_j = \frac{1}{2} (e_i e_j - e_j e_i) = e_i e_j$.
An element $X$ in $Cl_{p,q} (\mathbb{R})$ is expanded 
as
\begin{align}
X = x 1 + x^i e_i + \frac{1}{2!} x^{ij} e_i e_j + \frac{1}{3!} x^{ijk}
 e_i e_j e_k + \cdots + \frac{1}{n!} x^{1,\ldots, n} e_1 \cdots e_n,
\end{align}
where $x, x^i, x^{ij}, \ldots \in \mathbb{R}$ are coefficients.
The dimension of the algebra is 
\begin{align}
\dim (Cl_{p,q} (\mathbb{R})) = 1 + {}_n C_1 + {}_n C_2 + \cdots + {}_n C_n = 2^n.
\end{align}
The algebra involves $2^{n-1}$ imaginary and real units and $Cl_{p,q}
 (\mathbb{R})$ are in general non-commutative unital associative division algebras.
\paragraph{1-dim.}
The $2^0 = 1$-dimensional algebra is $Cl_{0,0} (\mathbb{R})$ only.
This is generated by $\{1\}$ and identified with $\mathbb{R}$, $Cl_{0,0}
(\mathbb{R}) \simeq \mathbb{R}$.

\paragraph{2-dim.}
The $2^1 = 2$-dimensional algebras are $Cl_{1,0} (\mathbb{R})$ and $Cl_{0,1} (\mathbb{R})$.
The algebra $Cl_{1,0} (\mathbb{R})$ is generated by $1$ and $e_1^2 = 1$
and they commute. Then this is isomorphic to the split-complex numbers
$Cl_{1,0} (\mathbb{R}) \simeq \mathrm{Sp}\mathbb{C}$.
The algebra $Cl_{0,1} (\mathbb{R})$ is generated by $1$ and $e_1^2 = -1$
and they commute. It is obvious that this is equivalent to the complex numbers
$Cl_{0,1} (\mathbb{R}) \simeq \mathbb{C}$.

\paragraph{4-dim.}
The $2^2 = 4$-dimensional algebras are $Cl_{2,0} (\mathbb{R})$,
$Cl_{1,1}(\mathbb{R})$ and $Cl_{0,2} (\mathbb{R})$.
The algebra $Cl_{2,0} (\mathbb{R})$ is generated by $1$, $e_1, e_2$ and
$e_3 = e_1 e_2$. Since $e_1$ and $e_2$ anti-commute $\{e_1, e_2\} = 0$, we have
\begin{gather}
\{e_1, e_2\} = \{e_2, e_3\} = \{e_3, e_1\} = 0,
\notag \\
e_1^2 = e_2^2 = 1, \qquad e_3^2 = -1,
\notag \\
e_1 e_2 = e_3, \qquad e_2 e_3 = - e_1, \qquad e_3 e_1 = - e_2.
\end{gather}
This is equivalent to the algebra of the split-quaternions
$Cl_{2,0} (\mathbb{R}) \simeq \mathrm{Sp}\mathbb{H}$.

The algebra $Cl_{1,1} (\mathbb{R})$ is generated by $1$, $e_1$,
$e_2$ and $e_3 = e_1 e_2$.
They satisfy
\begin{gather}
\{e_1, e_2\} = \{e_2, e_3\} = \{e_3, e_1\} = 0,
\notag \\
e_1^2 = 1, \qquad e_2^2 = -1, \qquad e_3^2 = 1,
\notag \\
e_1 e_2 = e_3, \qquad e_2 e_3 = e_1, \qquad e_3 e_1 = - e_2.
\end{gather}
This is again equivalent to the split-quaternions 
$Cl_{1,1} (\mathbb{R}) \simeq \mathrm{Sp} \mathbb{H}$.

The algebra $Cl_{0,2} (\mathbb{R})$ is generated by $1$, $e_1, e_2$, and
$e_3 = e_1 e_2$.
They satisfy
\begin{gather}
\{e_1, e_2 \} = \{ e_2, e_3 \} = \{ e_3, e_1 \} = 0,
\notag \\
e_1^2 = e_2^2 = e_3^2 = -1,
\notag \\
e_1 e_2 = e_3, \qquad e_2 e_3 = e_1, \qquad e_3 e_1 = e_2.
\end{gather}
This is nothing but the algebra of quaternions 
$Cl_{0,2} (\mathbb{R}) \simeq \mathbb{H}$.

\paragraph{8-dim.}
The $2^3 = 8$-dimensional algebras are $Cl_{3,0} (\mathbb{R})$, 
$Cl_{2,1} (\mathbb{R})$,
$Cl_{1,2} (\mathbb{R})$ and $Cl_{0,3} (\mathbb{R})$.

The algebra $Cl_{3,0} (\mathbb{R})$ is generated by $1$, $e_1, e_2, e_3$ 
satisfying $e_1^2 = e_2^2 = e_3^2 = 1$ and
\begin{align}
e_4 = e_1 e_2, \qquad 
e_5 = e_2 e_3, \qquad
e_6 = e_3 e_1, \qquad
e_7 = e_1 e_2 e_3.
\label{eq:4dim_basis}
\end{align}
Since $\{e_i, e_j\} = 0, \ (i,j=1,2,3)$, we have
\begin{align}
e_4^2 = e_5^2 = e_6^2 = e_7^2 = -1.
\end{align}
We also have
\begin{gather}
\begin{aligned}
\{e_4, e_5\} &= e_1 e_2 e_2 e_3 + e_2 e_3 e_1 e_2 = e_1 e_3 + e_3 e_1 =
 0,
\notag \\
\{e_5, e_6\} &= e_2 e_3 e_3 e_1 + e_3 e_1 e_2 e_3 = e_2 e_1 + e_1 e_2 =
 0,
\notag \\
\{e_6, e_4\} &= e_3 e_1 e_1 e_2 + e_1 e_2 e_3 e_1 = e_3 e_2 + e_2 e_3 =
 0,
\notag 
\end{aligned} \\
\begin{align}
e_4 e_5 &= e_1 e_2 e_2 e_3 = e_1 e_3 = - e_6,
\notag \\
e_5 e_6 &= e_2 e_3 e_3 e_1 = e_2 e_1 = - e_4,
\notag \\
e_6 e_4 &= e_3 e_1 e_1 e2 = e_3 e_2 = - e_5.
\end{align}
\end{gather}
This is the algebra of the bi-quaternions 
$Cl_{3,0} (\mathbb{R}) \simeq \mathbb{C} \otimes \mathbb{H}$.

The algebra of $Cl_{2,1} (\mathbb{R})$ is generated by $1$, $e_1, e_2,
e_3$ satisfying $e_1^2 = e_2^2 = -1$, $e_3^2 = 1$
and \eqref{eq:4dim_basis}.
Since $\{e_i,e_j\} = 0, \ (i,j=1,2,3)$, we have
\begin{align}
e_4^2 = e_7^2 = -1, \qquad e_5^2 = e_6^2 = 1.
\end{align}
We find that $Cl_{2,1} (\mathbb{R})$ is again isomorphic to 
the bi-quaternions $Cl_{2,1} (\mathbb{R}) \simeq \mathbb{C} \otimes \mathbb{H}$.
Note that $(e_1, e_2, e_4)$ defines the quaternion subalgebra;
\begin{gather}
\{e_1, e_2\} = \{e_2, e_4\} = \{e_4,e_1\} = 0,
\notag \\
e_1 e_2 = e_4, 
\qquad 
e_2 e_4 = e_1,
\qquad 
e_4 e_1 = e_2.
\end{gather}

The algebra $Cl_{1,2} (\mathbb{R})$ is generated by $1$, $e_1, e_2, e_3$
satisfying $e_1^2 = 1, e_2^2 = e_3^2 = -1$ 
and \eqref{eq:4dim_basis}.
Since $\{e_i,e_j\} = 0, \ (i,j=1,2,3)$, we have
\begin{align}
e_4^2 = e_6^2 = 1, \qquad 
e_5^2 = e_7^2 = -1.
\end{align}
We find that $Cl_{1,2} (\mathbb{R})$ is isomorphic to bi-quaternions;
$Cl_{1,2} (\mathbb{R}) \simeq \mathbb{C} \otimes \mathbb{H}$.
Note that $(e_2,e_3,e_5)$ forms the quaternion subalgebra.

The algebra $Cl_{0,3}(\mathbb{R})$ is generated by $1$, $e_1, e_2, e_3$ 
satisfying $e_1^2 = e_2^2 = e_3^2 = -1$ 
and \eqref{eq:4dim_basis}.
Since $\{e_i,e_j\} = 0, \ (i,j=1,2,3)$ we have
\begin{align}
e_1^2 = e_2^2 = e_3^2 = -1, \qquad 
e_4^2 = e_5^2 = e_6^2 = -1, \qquad e_7^2 = 1.
\end{align}
The basis $e_4,e_5,e_6$ all anti-commute.
This is 
the split-bi-quaternions
$Cl_{0,3} (\mathbb{R}) \simeq \mathrm{Sp} \mathbb{C} \otimes \mathbb{H}$.

\paragraph{16-dim.}
The $2^4 = 16$-dimensional algebras are $Cl_{4,0} (\mathbb{R})$, $Cl_{3,1} (\mathbb{R})$,
$Cl_{2,2} (\mathbb{R})$, $Cl_{1,3} (\mathbb{R})$ and $Cl_{0,4} (\mathbb{R})$.

The algebra $Cl_{4,0} (\mathbb{R})$ is generated by $1, e_1, e_2, e_3, e_4$
satisfying $e_1^2 = e_2^2 = e_3^2 = e_4^2 = 1$ and
\begin{align}
e_5 &= e_1 e_2,
\quad
e_6 = e_1 e_3,
\quad 
e_7 = e_1 e_4,
\quad
e_8 = e_2 e_3,
\quad
e_9 = e_2 e_4,
\quad
e_{10} = e_3 e_4,
\notag \\
e_{11} &= e_1 e_2 e_3,
\quad
e_{12} = e_1 e_2 e_4,
\quad
e_{13} = e_1 e_3 e_4,
\quad
e_{14} = e_2 e_3 e_4,
\quad
e_{15} = e_1 e_2 e_3 e_4.
\label{eq:Clifford16d_components}
\end{align}
Since $\{e_i, e_j\} = 0, \ (i,j=1,2,3,4)$, we have
\begin{gather}
1^2 = e_1^2 = e_2^2 =e_3^2 = e_4^2 = e_{15}^2 = 1,
\notag \\
e_5^2 = \cdots = e_{14}^2 = -1.
\end{gather}
The algebra $Cl_{4,0} (\mathbb{R})$ contains 10 imaginary and 6 real units
and isomorphic to 
the split-tetra-quaternions
$Cl_{4,0} (\mathbb{R}) \simeq \mathrm{Sp} \mathbb{H} \otimes \mathbb{H}$.
Indeed, we can extract 10 quaternions as subalgebras;
\begin{alignat}{5}
&(e_5, e_6, e_8),
&\quad
&(e_5, e_7, e_9),
&\quad
&(e_5, e_{13}, e_{14}),
&\quad
&(e_6, e_7, e_{10}),
&\quad
&(e_6, e_{12}, e_{14}),
\notag \\
&(e_7, e_{11}, e_{14}),
&\quad
&(e_8, e_9, e_{10}),
&\quad
&(e_8, e_{12}, e_{13}),
&\quad
&(e_9, e_{11}, e_{13}),
&\quad
&(e_{10}, e_{11}, e_{12}).
\end{alignat}
This is equivalent to the algebra \eqref{eq:basis_stq}.

The algebra $Cl_{3,1} (\mathbb{R})$ is generated by $1, e_1, e_2, e_3, e_4$
satisfying $e_1^2 = e_2^2 = e_3^2 = 1$, $e_4^2 = -1$ and
\eqref{eq:Clifford16d_components}.
They satisfy
\begin{gather}
1^2 = e_1^2 = e_2^2 = e_3^2 = e_7^2 = e_9^2 = e_{10}^2 = e_{12}^2 =
 e_{13}^2 = e_{14}^2 = 1,
\notag \\
e_4^2 = e_5^2 = e_6^2 = e_8^2 = e_{11}^2 = e_{15}^2 = -1.
\end{gather}
This contains 6 imaginary and 10 real units. 
This is isomorphic to 
the tetra-quaternions
$Cl_{3,1} (\mathbb{R}) \simeq \mathbb{H} \otimes \mathbb{H}$.

The algebra $Cl_{2,2} (\mathbb{R})$ is generated by $1, e_1, e_2, e_3, e_4$,
satisfying $e_1^2 = e_2^2 = 1$, $e_3^2 = e_4^2 = -1$ 
and \eqref{eq:Clifford16d_components}. They define
6 imaginary and 10 real units;
\begin{gather}
1^2 = e_1^2 = e_2^2 = e_6^2 = e_7^2 = e_8^2 = e_9^2 = e_{11}^2 =
 e_{12}^2 = e_{15}^2 = 1,
\notag \\
e_3^2 = e_4^2 = e_5^2 = e_{10}^2 = e_{13}^2 = e_{14}^2 = -1.
\end{gather}
The algebra is isomorphic to 
the tetra-quaternions
$Cl_{2,2} (\mathbb{R}) \simeq \mathbb{H} \otimes \mathbb{H}$.

The algebra $Cl_{1,3} (\mathbb{R})$ is generated by $1, e_1, e_2, e_3$
satisfying $e_1^2 = 1$, $e_2^2 = e_3^2 = e_4^2 = -1$
and \eqref{eq:Clifford16d_components}. They define 10 imaginary and 6
real units;
\begin{gather}
1^2 = e_1^2 = e_5^2 = e_6^2 = e_7^2 = e_{14}^2 = 1,
\notag \\
e_2^2 = e_3^2 = e_4^2 = e_8^2 = e_9^2 = e_{10}^2 = e_{11}^2 = e_{12}^2 =
 e_{13}^2 = e_{15}^2 = -1.
\end{gather}
The algebra is isomorphic to 
the split-tetra-quaternions 
$Cl_{1,3} (\mathbb{R}) \simeq \mathrm{Sp} \mathbb{H} \otimes \mathbb{H}$.

Finally, the algebra $Cl_{0,4}(\mathbb{R})$ is generated by $1, e_1,
e_2, e_3, e_4$ satisfying $e_1^2 = e_2^2 = e_3^2 = e_4^2 = -1$
and \eqref{eq:Clifford16d_components}. They define 10 imaginary and 6
real units;
\begin{gather}
1^2 = e_{11}^2 = e_{12}^2 = e_{13}^2 = e_{14}^2 = e_{15}^2 = 1,
\notag \\
e_1^2 = e_2^2 = e_3^2 = e_4^2 = e_5^2 = e_6^2 = e_7^2 = e_8^2 = e_9^2 =
 e_{10}^2 = -1.
\end{gather}
The algebra is isomorphic to 
the split-tetra-quaternions
$Cl_{0,4} (\mathbb{R}) \simeq \mathrm{Sp} \mathbb{H} \otimes \mathbb{H}$.

\paragraph{32-dim.}
The $2^5 = 32$-dimensional algebras are 
$Cl_{5,0} (\mathbb{R})$,
$Cl_{4,1} (\mathbb{R})$,
$Cl_{3,2} (\mathbb{R})$,
$Cl_{2,3} (\mathbb{R})$,
$Cl_{1,4} (\mathbb{R})$ and
$Cl_{0,5} (\mathbb{R})$.
For example, $Cl_{5,0} (\mathbb{R})$ is generated by the basis
\begin{align}
+1 &: \ 
1, e_1, e_2, e_3, e_4, e_5,
\notag \\
-1 &: \ 
e_1 e_2, e_1 e_3, e_1 e_4, e_1 e_5,
e_2 e_3, e_2 e_4, e_2 e_5,
e_3 e_4, e_3 e_5,
e_4 e_5,
\notag \\
-1 &: \
e_1 e_2 e_3, e_1 e_2 e_4, e_1 e_2 e_5,
e_1 e_3 e_4, e_1 e_3 e_5, e_1 e_4 e_5,
e_2 e_3 e_4, e_2 e_3 e_5, e_2 e_4 e_5,
e_3 e_4 e_5,
\notag \\
+1 &: \ 
e_1 e_2 e_3 e_4, e_1 e_2 e_3 e_5, e_1 e_2 e_4 e_5, e_1 e_3 e_4 e_5, e_2
 e_3 e_4 e_5,
\notag \\
+1 &: \
e_1 e_2 e_3 e_4 e_5.
\label{eq:32_basis}
\end{align}
Here $+1$ and $-1$ stand for the real and imaginary units.
Therefore $Cl_{5,0} (\mathbb{R})$ involves 12 real and 20 imaginary units.
We show only the numbers of real and imaginary units 
of the other Clifford algebras;
\begin{quote}
\begin{itemize}
\item $Cl_{5,0} (\mathbb{R})$ : 12 real and 20 imaginary,
\item $Cl_{4,1} (\mathbb{R})$ : 16 real and 16 imaginary,
\item $Cl_{3,2} (\mathbb{R})$ : 20 real and 12 imaginary,
\item $Cl_{2,3} (\mathbb{R})$ : 16 real and 16 imaginary,
\item $Cl_{1,4} (\mathbb{R})$ : 12 real and 20 imaginary,
\item $Cl_{0,5} (\mathbb{R})$ : 16 real and 16 imaginary.
\end{itemize}
\end{quote}
We find that $Cl_{3,2} (\mathbb{R})$ is isomorphic to 
the
split-bi-quaternions over $\mathbb{H}$ which has 20 real and 12 imaginary
units $Cl_{3,2} (\mathbb{R}) \simeq \mathrm{Sp} \mathbb{C} \otimes
\mathbb{H} \otimes \mathbb{H}$.

\paragraph{64-dim.}
The $2^6 = 64$-dimensional algebra contains
$Cl_{6,0} (\mathbb{R})$,
$Cl_{5,1} (\mathbb{R})$,
$Cl_{4,2} (\mathbb{R})$,
$Cl_{3,3} (\mathbb{R})$,
$Cl_{2,4} (\mathbb{R})$,
$Cl_{1,5} (\mathbb{R})$ and
$Cl_{0,6} (\mathbb{R})$.
We can show that 
the split-tetra-quaternions over $\mathbb{H}$, $\mathrm{Sp}\mathbb{H} \otimes
\mathbb{H} \otimes \mathbb{H}$ is involved in $Cl_{p,q} (\mathbb{R})$.

\paragraph{Clifford algebra over $\mathbb{C}$.}
As is clear from the construction, 
the Clifford algebra over the field $\mathbb{R}$ always 
has an anti-commutative basis when the dimension is greater than 
or equal to four. 
Therefore, the four-dimensional algebra of the bi-complex numbers, that
consists of commuting basis, cannot be written in Clifford algebras.
This is not the case when the field defining Clifford 
algebras is changed from $\mathbb{R}$ to $\mathbb{C}$.
For example, $Cl_0 (\mathbb{C})$ is a complex vector space generated by
$1$. This is identified with $\mathbb{C}$.
The complex 2-dimensional (hence the real 4-dimensional) algebra $Cl_1 (\mathbb{C})$ is generated by
$1$ and $e_1$ satisfying $e_1^2 = 1$, i.e.,
\begin{align}
Z = z_1 1 + z_2 e_1, \qquad z_1, z_2 \in \mathbb{C}.
\end{align}
In terms of the real basis, this is generated by 
\begin{align}
1, \quad i, \quad e_1, \quad i e_1.
\end{align}
Note that they all commute and define two real and two imaginary units;
\begin{align}
1^2 = e_1^2 = 1, \qquad i^2 = (ie_1)^2 = -1.
\end{align}
It is obvious that this is equivalent to the algebra of the bi-complex numbers,
$Cl_1 (\mathbb{C}) \simeq \mathbb{C} \otimes \mathbb{C}$.
In the same way, we have isomorphisms $Cl_2 (\mathbb{C}) \simeq
\mathbb{C} \otimes \mathbb{H}$, $Cl_3 (\mathbb{C}) \simeq \mathbb{C}
\otimes \mathbb{C} \otimes \mathbb{H}$, and so on.

We note that not all the hypercomplex numbers are isomorphic to Clifford algebras.
For example, the tri-complex numbers by Segre $\mathbb{C}_3 = \mathbb{C}
\otimes \mathbb{C} \otimes \mathbb{C}$ is not obtained in this way.
A summary of the algebras is found in Table \ref{tb:Clifford}.
\begin{table}[t]
\centering
\begin{tabular}{c|c|c|l}
dim & Clifford algebras & hypercomplex numbers &  
\\
\hline
1 & $Cl_{0,0} (\mathbb{R})$ & $\mathbb{R}$ & real numbers
\\
\hline
2 & $Cl_{1,0} (\mathbb{R})$ & $\mathrm{Sp}\mathbb{C}$ & split-complex numbers
\\
2 & $Cl_{0,1} (\mathbb{R})$ & $\mathbb{C}$ & complex numbers
\\
\hline
4 & $Cl_{2,0} (\mathbb{R})$ & $\mathrm{Sp} \mathbb{H}$ & split-quaternions
\\
4 & $Cl_{1,1} (\mathbb{R})$ & $\mathrm{Sp} \mathbb{H}$ & split-quaternions
\\
4 & $Cl_{0,2} (\mathbb{R})$ & $\mathbb{H}$ & quaternions
\\
\hline
8 & $Cl_{3,0} (\mathbb{R})$ & $\mathbb{C} \otimes \mathbb{H}$ & bi-quaternions
\\
8 & $Cl_{2,1} (\mathbb{R})$ & $\mathbb{C} \otimes \mathbb{H}$ & bi-quaternions
\\
8 & $Cl_{1,2} (\mathbb{R})$ & $\mathbb{C} \otimes \mathbb{H}$ & bi-quaternions
\\
8 & $Cl_{0,3} (\mathbb{R})$ & $\mathrm{Sp} \mathbb{C} \otimes
	 \mathbb{H}$ & split-bi-quaternions
\\
\hline
16 & $Cl_{4,0} (\mathbb{R})$ & $\mathrm{Sp} \mathbb{H} \otimes
	 \mathbb{H}$ & split-tetra-quaternions
\\
16 & $Cl_{3,1} (\mathbb{R})$ & $\mathbb{H} \otimes \mathbb{H}$ & tetra-quaternions
\\
16 & $Cl_{2,2} (\mathbb{R})$ & $\mathbb{H} \otimes \mathbb{H}$ & tetra-quaternions
\\
16 & $Cl_{1,3} (\mathbb{R})$ & $\mathrm{Sp} \mathbb{H} \otimes
	 \mathbb{H}$ & split-tetra-quaternions
\\
16 & $Cl_{0,4} (\mathbb{R})$ & $\mathrm{Sp} \mathbb{H} \otimes
	 \mathbb{H}$ & split-tetra-quaternions
\\
\hline
\hline
2 & $Cl_0 (\mathbb{C})$ & $\mathbb{C}$ & complex numbers
\\
\hline
4 & $Cl_1 (\mathbb{C})$ & $\mathbb{C} \otimes \mathbb{C}$ & bi-complex numbers
\\
\hline
8 & $Cl_2 (\mathbb{C})$ & $\mathbb{C} \otimes \mathbb{H}$ & bi-quaternions
\\
\hline
16 & $Cl_3 (\mathbb{C})$ & $\mathbb{C} \otimes \mathbb{C} \otimes
	 \mathbb{H}$ & bi-quaternions over $\mathbb{C}$
\end{tabular}
\caption{Clifford algebra and hypercomplex numbers.
}
\label{tb:Clifford}
\end{table}

\end{appendix}


}
\end{document}